# Semantic Units: Organizing knowledge graphs into semantically meaningful units of representation

Vogt, Lars[1]; Kuhn, Tobias[2]; Hoehndorf, Robert[3]

[1] *TIB Leibniz Information Centre for Science and Technology, Welfengarten 1B, 30167 Hanover, Germany*, [orcid.org/0000-0002-8280-0487](https://orcid.org/0000-0002-8280-0487)

[2] *Department of Computer Science, Vrije Universiteit Amsterdam, Netherlands,* [orcid.org/0000-0002-1267-0234](https://orcid.org/0000-0002-1267-0234)

[3] *Computational Bioscience Research Center, Computer, Electrical and Mathematical Sciences & Engineering Division, King Abdullah University of Science and Technology, 4700 KAUST, 23955 Thuwal, Saudi Arabia,* [orcid.org/0000-0001-8149-5890](https://orcid.org/0000-0001-8149-5890)

Correspondence to [lars.m.vogt@googlemail.com](mailto:lars.m.vogt@googlemail.com)

# Abstract

**Background:** Knowledge graphs and ontologies are becoming increasingly important as technical solutions for Findable, Accessible, Interoperable, and Reusable data and metadata (FAIR Guiding Principles). We discuss four challenges that impede the use of FAIR knowledge graphs.

**Results:** Semantic units have the potential to solve the challenges by structuring a knowledge graph into identifiable and semantically meaningful subgraphs. Each semantic unit is represented by its own resource, instantiates a corresponding semantic unit class, and can be implemented as a FAIR Digital Object and a nanopublication in RDF/OWL and property graphs. We distinguish statement and compound units as basic categories of semantic units. Statement units represent smallest, independent propositions that are semantically meaningful for a human reader. They consist of one or more triples and mathematically partition a knowledge graph. We distinguish assertional, contingent (prototypical), and universal statement units as basic types of statement units and propose representational schemes and formal semantics for them (including for absence statements, negations, and cardinality restrictions) that do not involve blank nodes and that translate back to OWL. Compound units, on the other hand, represent semantically meaningful collections of semantic units and we distinguish various types of compound units, representing different levels of representational granularity, different types of granularity trees, and different frames of reference.

**Conclusions:** Semantic units support making statements about statements, can be used for graph-alignment, subgraph-matching, knowledge graph profiling, and for managing access restrictions to sensitive data. Organizing the graph into semantic units supports the separation of ontological, diagnostic (i.e., referential), and discursive information, and it also supports the differentiation of multiple frames of reference.

# Background

In times of ever-increasing amounts of data being created every day (1–3), new technical and societal challenges arise (4) that ask for innovative ways of representing and managing data in science and industry. Being able to collect, integrate, and analyze large amounts of data from various sources also represents one of the requirements for facing biodiversity loss and climate change, two major global challenges we are currently facing (5). Solutions to these problems will be driven by sharing data across many stakeholders, needing an effort to help interlink data providers from a diverse range of different areas, often requiring a truly interdisciplinary approach (6), in which data stewardship remains in the hands of the domain experts or institutions, thus ensuring their technical autonomy (following Barend Mons' *data visiting* as opposed to *data sharing* (7)).

From a data management and data representation perspective, this requires data and metadata to be **FAIR**, i.e., readily **F**indable, **A**ccessible, **I**nteroperable, and **R**eusable for **machines and humans alike** (8). If this is not the case, Big Data ultimately turns into Dark Data (9). The establishment of the FAIR Guiding Principles as a general standard in science and industry would also contribute to a solution for the reproducibility crisis in science (10) and the question of the trustworthiness of information in general (see also **TRUST** Principles of **T**ransparency, **R**esponsibility, **U**ser Focus, **S**ustainability, and **T**echnology (11)). Therefore, we must build something along the lines of the Internet of FAIR Data and Services (12) that scales with Big Data, through which all relevant data-rich institutions, research projects, and citizen-science projects can make their data and metadata accessible following the FAIR Guiding Principles (13,14). This requires providing rich machine-actionable data and metadata with human-readable interface outputs and search capabilities, and organizing this data into **FAIR Digital Objects** (15,16), each of which possesses its own Unique Persistent and Resolvable Identifier (UPRI) for referencing it individually.

In this context, **knowledge graphs** can substantially contribute to the needed technical solutions, providing a suitable framework for managing and representing FAIR data and metadata (17). Knowledge graphs are becoming increasingly popular (18), especially after the 2012 announcement of the Google Knowledge Graph (19) which was followed by further announcements of knowledge graphs being developed from industry and by a growing number of scientific publications on knowledge graphs (20). Besides general applications in industry and research, knowledge graphs are thereby particularly applied in the context of semantic search based on entities and relations, deep reasoning, disambiguation of natural language, machine reading, and entity consolidation for Big Data and text analytics (21).

The graph-based abstractions employed in knowledge graphs have several benefits compared to relational or other NoSQL models, including (i) an intuitive way for modelling relations, (ii) allowing postponing specifications of definitions for data schema so that they can flexibly evolve, which is especially important when dealing with incomplete knowledge, (iii) employing machine-actionable knowledge representation formalisms such as ontologies and rules, (iv) applying graph analytics and machine learning, and (v) utilizing the specialized graph query languages of knowledge graphs that support, in addition to standard relational operators such as joins, unions, and projections, also navigational operators for recursively searching for entities through arbitrary-length paths (20,22–27). Moreover, due to their inherent semantic transparency, knowledge graphs can improve the transparency of data-based decision-making and improve communication in research and science in general.

However, although providing a suitable technical framework, using a knowledge graph for documenting data and metadata does not *necessarily* result in FAIR data and metadata, but requires following specific guidelines such as consistently applying adequate semantic data models and organizing data into FAIR Digital Objects (15,16). Moreover, as it is often the case with new technologies, knowledge graphs bring their own specific technical, conceptual, and societal challenges. This already begins with the concept of a knowledge graph, which is somewhat fuzzy (20) and covers different technical and conceptual incarnations, including property graphs such as Neo4J ([https://neo4j.com/](https://neo4j.com/)) and approaches based on the Resource Description Framework (RDF), the use of RDF-stores, and, with the Web Ontology Language (OWL), also applications of Description Logics.

Here, we first briefly discuss four of these challenges, and then we introduce the idea of partitioning and structuring a knowledge graph into **identifiable and semantically meaningful units of representation (short: semantic units)**. The concept of semantic units can significantly contribute to solutions for the four challenges. We introduce the two basic categories of semantic units, i.e., statement units and compound units, as new elements in FAIR knowledge graphs in addition to the well-known triples and the graph as a whole. Statement and compound units can be employed to organize the data graph into five levels of representational granularity, ranging from the level of individual triples to the level of the graph as a whole. We introduce additional subcategories of semantic units that can be used to further organize the data graph. We continue arguing that because each semantic unit can be organized as a FAIR Digital Object that possesses its own UPRI, semantic units can be referred to within triple statements, thus providing a very efficient way of making statements about statements. With the introduction of semantic units, we follow a user-centric approach and **add another layer of triples on top of the well established RDF and OWL layer for knowledge graphs** (Fig. 1). By simplifying the semantic modelling of empirical data and reducing their representational complexity and by providing representations (i.e., patterns) and formal semantics for statements for which in OWL no formal semantics exist, semantic units increase the usability of knowledge graphs for domain-experts and developers alike.

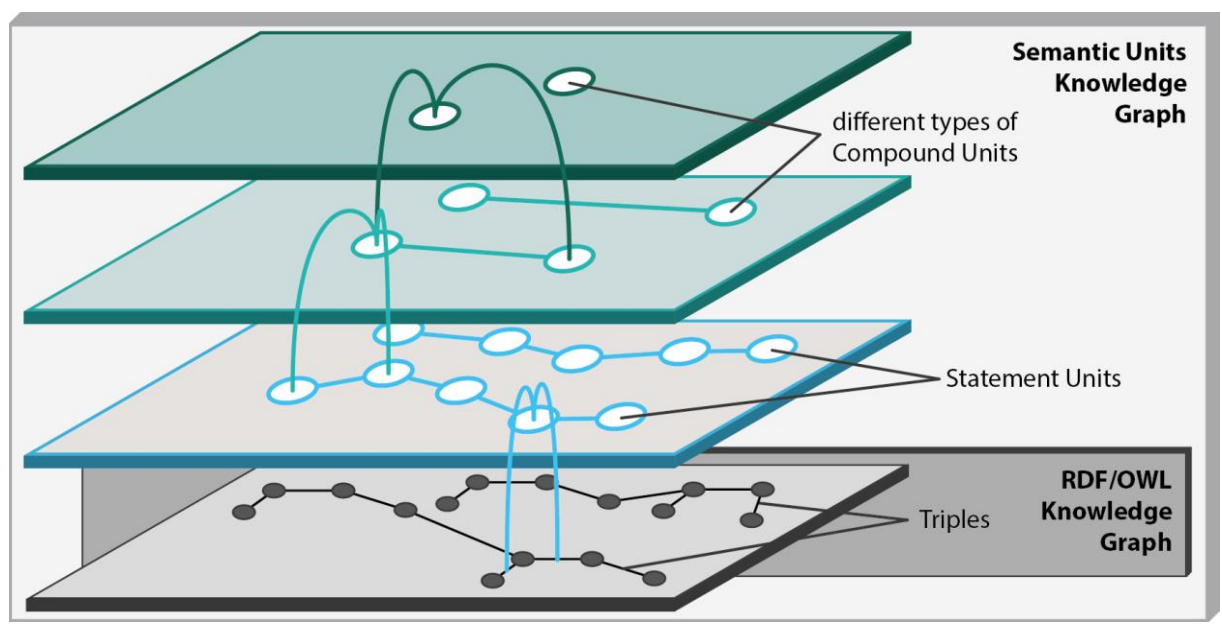

**Figure 1: Semantic units add further layers to a knowledge graph on top of the RDF/OWL layer of triples.** The layer of triples is mathematically partitioned into a layer of statement units, so that each triple belongs to exactly one statement unit and each statement unit comprises one or more triples. Statement units can be organized into different types of semantically meaningful collections (i.e., compound units) with which various additional layers can be defined to further structure and organize the knowledge graph in semantically meaningful ways.

> **Box 1 | Conventions**
>
> In this paper, we refer to FAIR knowledge graphs as machine-actionable semantic graphs for documenting, organizing, and representing assertional (e.g., empirical data), universal, and contingent statements and thus a mixture of ABox and TBox expressions (thereby contrasting knowledge graphs with ontologies, with the latter containing mainly universal statements and thus TBox expressions). We want to point out that we discuss semantic units against the background of RDF-based triple stores, OWL, and Description Logics as a formal framework for inferencing, and labeled property graphs as an alternative to triple stores, because these are the main technologies and logical frameworks used in knowledge graphs that are supported by a broad community of users and developers and for which accepted standards exist. We are aware of the fact that alternative technologies and frameworks exist that support an n-tuples syntax and more advanced logics (e.g., First Order Logic) (28,29), but supporting tools and applications are missing or are not widely used to turn them into well-supported, scalable, and easily usable knowledge graph applications.
>
> Throughout this text we use regular underlined to indicate ontology classes, *italicsUnderlined* when referring to properties (i.e., relations in Neo4j), and use ID numbers to specify each. ID numbers are composed of the ontology prefix followed by a colon and a number, e.g., *isAbout* (IAO:0000136). If the term is not yet covered in any ontology, we indicate it with *, e.g., the class *metric measurement statement unit*. We use 'regular underlined' to indicate instances of classes,

> with the label referring to the class label and the ID number to the class. Moreover, when we use the term *resource*, we understand it to be something that is uniquely designated (e.g., a Uniform Resource Identifier, URI) and about which you want to say something. It thus stands for something and represents something you want to talk about. In RDF, the *Subject* and the *Predicate* in a triple statement are always resources, whereas the *Object* can be either a resource or a literal. Resources can be either properties, instances, or classes, with properties taking the *Predicate* position in a triple and with instances referring to individuals (=particulars) and classes to universals.
>
> For reasons of clarity, in the text and in all figures, we represent resources not with their UPRIs but with human-readable labels, with the implicit assumption that every property, every instance, and every class has its own UPRI.

## Challenge 1: FAIR empirical data must specify the graph patterns used for their modelling to prevent schematic interoperability conflicts

FAIR is often understood to mean that for data and metadata statements to be interoperable and reusable, all concepts used in them must have identifiers, which in turn are provided by controlled vocabularies such as ontologies. What is frequently overlooked is the fact that to be FAIR, not only the concepts must be standardized (i.e., terminological interoperability), but also the way they are related to one another in data and metadata statements and thus the statements' underlying **semantic graph patterns** (i.e., schematic interoperability)—especially when stored and managed in a knowledge graph. In case of universal statements, in RDF-based knowledge graphs this graph pattern is typically well-defined using OWL-specific object properties in class axioms (see also Figure 5B), resulting in TBox expressions.

With empirical data, the situation is different. Empirical data should be modelled as ABox expressions (30). However, due to the high general expressivity of RDF and OWL, in a knowledge graph, any given empirical data statement can be modelled in many, usually not directly interoperable ways. A machine would have a hard time to identify two differently structured ABox expressions that actually model the same underlying data statement. As a result, we must deal with such schematic interoperability conflicts, whenever data are modelled using different graph patterns (cf. Fig. 2 and Fig. 3).

## Observation:
*ObjectX weighs 5 kilograms*

## Observation Graph:

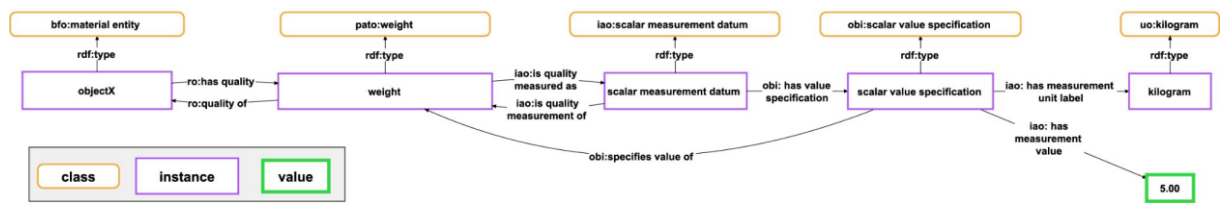

**Figure 2: Comparison of a human-readable statement with its machine-actionable representation as an ABox semantic graph following the RDF syntax. Top**: A human-readable statement about the observation that ObjectX weighs 5 kilograms. **Bottom**: A representation of the same statement as a graph, using RDF and following the general pattern for measurement data from the Ontology for Biomedical Investigations (OBI; http://obi-ontology.org/) (31) of the Open Biological and Biomedical Ontology Foundry (OBO; http://www.obofoundry.org/).

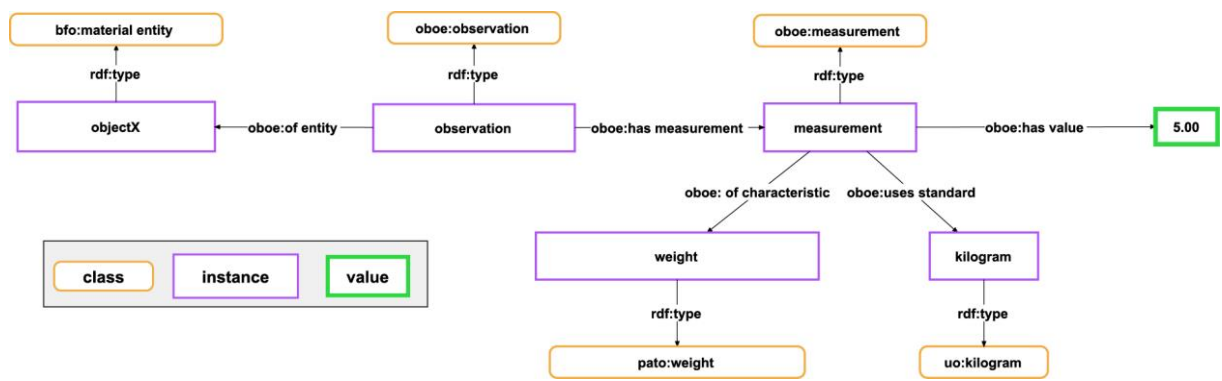

**Figure 3: Alternative machine-actionable representation of the data statement from Fig. 2, following the RDF syntax and the graph-model from the Extensible Observation Ontology (OBOE).** This graph represents the same data statement as shown in Figure 2 Top, but applies a different semantic graph model for its representation, which is based on OBOE (http://bioportal.bioontology.org/ontologies/OBOE), an ontology frequently used in the ecology community.

Therefore, for an ABox representation of an empirical data statement to be FAIR, one must know which graph pattern has been used for its semantic modelling. Only instance-based graphs (i.e., graphs with instance resources in the *Subject* and *Object* positions of their triples; ABoxes) that are modelled using the same template in the form of a graph pattern, for instance specified as a shape using SHACL (32), are guaranteed to meet the minimum requirement for interoperability. Ideally, statements of the same type, e.g., all weight measurements, use the same graph pattern to be potentially interoperable. Therefore, we need **identifiers for such graph patterns**, and if an empirical datum is documented in the form of an ABox, its metadata should reference the corresponding graph-pattern identifier. With this information, one can identify potentially interoperable ABox expressions by their commonly shared graph-pattern identifiers.

Practically, this implies (i) that all statements in a knowledge graph must be classified into statement classes, with each class having an associated graph pattern specified (e.g., in the form of a shape specification) and (ii) that the subgraph belonging to a particular statement must be identifiable. Only if these two criteria are met, ABox representations of data and metadata truly comply with the FAIR Guiding Principles. Semantic units provide a means to meet these two criteria.

## Challenge 2: Many software developers do not see the benefit of graph query languages

Most knowledge graphs are either directed labeled graphs that are based on RDF/OWL and stored in tuple stores, or they are labeled property graphs such as Neo4j. Directly interacting with these graphs, i.e., conducting CRUD operations for creating (=writing), reading (= searching), updating, and deleting statements in the knowledge graph, requires the use of a query language. For RDF/OWL, this is for example SPARQL ([https://www.w3.org/TR/rdf-sparql-query/](https://www.w3.org/TR/rdf-sparql-query/)), and for Neo4j, it is Cypher ([https://neo4j.com/developer/cypher/](https://neo4j.com/developer/cypher/)).

Whereas these query languages allow detailed and very complex queries, writing queries in SPARQL or Cypher is demanding. Users of knowledge graph applications usually lack the required background for writing such queries themselves. Unfortunately, our personal experiences are that even most developers are not familiar with these languages and struggle with their complexity when attempting to learn them. In other words, having to write SPARQL or Cypher queries represents an entry barrier for interacting with a knowledge graph and hinders their broader usage (33). We thus need technical solutions to mitigate this challenge, such as openly available and reusable query patterns that link to specific graph patterns.

The concept of semantic units can contribute to such solutions by allowing experts in Semantics to provide generic CRUD queries for each type of semantic unit that domain experts as users and developers of knowledge graphs that apply semantic units can employ without having to write the queries themselves. Queries of different types of semantic units can be combined via union or intersection to form more complex queries.

## Challenge 3: Making statements about statements can be challenging

The RDF triple syntax of *Subject*, *Predicate*, and *Object* does not natively provide a good method for making statements about statements. In RDF, a statement about a statement is a triple that relates a statement consisting of one or more triples to some value, resource, or some other such statement.

Unfortunately, if you wanted to represent empirical data or the contents of for instance a scholarly publication in a FAIR knowledge graph, you frequently have to model statements about statements. For example, if you want to provide detailed metadata for a measurement datum, you would actually have to relate two subgraphs to one another: the graph representing the measurement itself, documented in a set of triples (e.g., see Fig. 2) and the graph documenting the underlying measuring process, which also involves several triples (e.g., see Fig. 4).

In case we only want to refer to a single triple statement in another triple, RDF reification can be used. In RDF reification, a resource is defined to represent a particular triple by describing it via three additional triples that specify its *Subject*, *Predicate*, and *Object*. Alternatively, one can also use the RDF-star approach (34,35). In RDF-star, however, all statements involving the same subject, predicate, and object are understood to reference the same triple and are thus not distinguished, with the consequence that one cannot distinguish different instances of the same statement. Moreover, while using RDF-star is feasible for referring to a single triple, it becomes very inefficient and complicated to query when having to refer to sets of multiple triple statements and thus larger subgraphs. If you wanted to make a statement about the entire graph as depicted in Figure 4 using reification, you would first have to specify a statement for each triple in the graph. Since the graph comprises more than 20 triples, you would end up having to create more than 60 triples in this first step. However, because you want to refer to the graph as a whole and not to each of its triples individually, you would have to create more than 20 additional triples that link the resource representing the entire subgraph with the resources you created in the first step for representing each individual triple. While this is technically feasible, it is neither elegant nor easily queried.

In cases like this, using Named Graphs is much more efficient than applying RDF reification or RDF-star. Named Graphs can be used in RDF-based knowledge graphs. A Named Graph resource identifies a set of triple statements by adding the UPRI of the subgraph and thus the statement as a fourth element to each triple, turning the triples into quads. Moreover, Named Graphs have the additional benefit to outperform other metadata representation models when conducting more complex queries (36).

In labeled property graphs, on the other hand, assigning a resource for identifying subgraphs within the overall data graph is straightforward and can be achieved by adding the resource identifier as the value of a respective property-value pair and add this value pair to all relations and nodes that belong to the same subgraph.

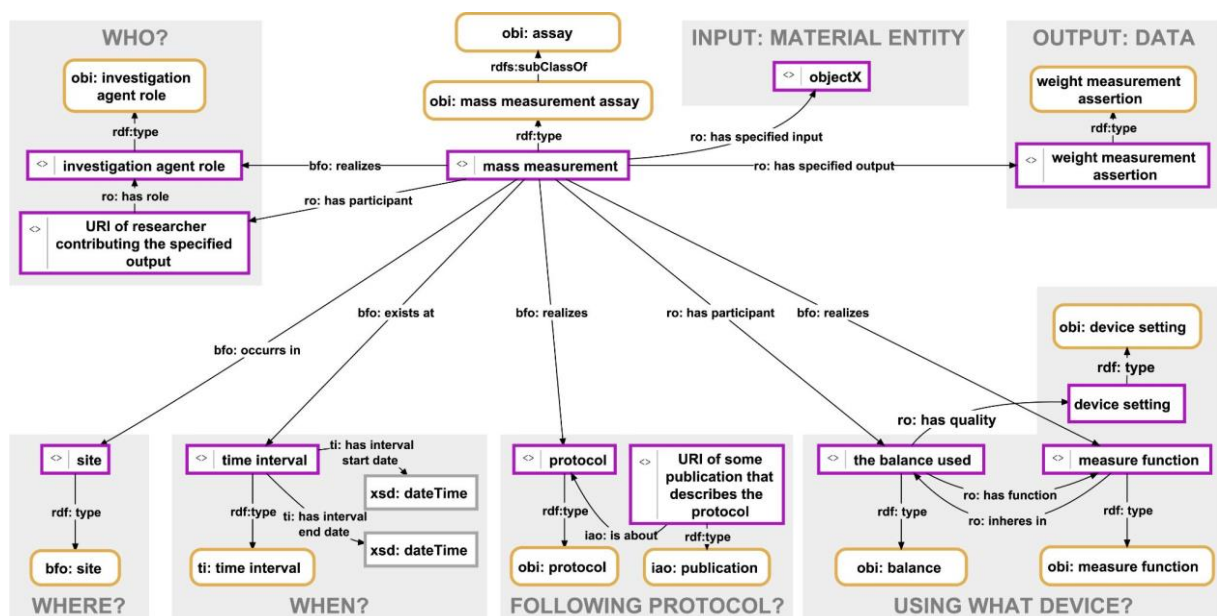

**Figure 4: A detailed machine-actionable representation of the metadata relating to a weight measurement datum documented as an RDF ABox graph.** The representation takes the form of an ABox semantic graph following the RDF syntax. The graph documents a mass measurement process using a balance. It relates an instance of mass measurement assay (OBI:0000445) with instances of various other classes from different ontologies, specifying who conducted the measurement, where and when it took place, following which protocol and using which device (i.e., balance). The graph furthermore specifies the particular material entity that served as subject and thus as input of the measurement process (i.e., '*objectX*'), and it specifies the data that is the output of the observation, which is contained in a particular weight measurement assertion.

Organizing the overall data graph into different semantic units and thus semantically meaningful subgraphs at different levels of representational granularity provides an efficient way of structuring the graph to allow users to intuitively make statements about statements, and also provides a clear and straightforward implementation schema that is beneficial for developing relevant search schemata.

# Challenge 4: Conceptual gap between RDF/OWL and property graphs and problems of distinguishing universal, contingent, and assertional statements

Some knowledge graphs are based on RDF/OWL, others on labeled property graphs. Each technology has its specific advantages and shortcomings (see table 1).

**Table 1: Comparison of the advantages and shortcomings of knowledge graphs based on RDF/OWL and on labeled property graphs** (37,38).

|  | RDF | labeled property graph |
|---|---|---|
| query language | SPARQL, W3C recommendation | various, none has a W3C recommendation |
| formal semantics and reasoning | OWL; however:<br>1. when mapped to RDF, OWL is usually very verbose, leading to unnecessarily complex graphs that are not intuitively comprehensible for a human reader and inhibit effective computation;<br>2. OWL is not the only mode of inference, and other frameworks besides Description Logics exist for inferencing | no unified set of standards (no W3C recommendation)—a standardized mapping of OWL to a property graph is still lacking; the data model of Neo4j is not based on a formal set theory or first-order-logic |
| relating information directly to edges | only indirectly through reification, RDF-star, or named graphs (see *Challenge 3*) | information can be directly related to relations, i.e., edges |
| distinction of universal, contingent, and assertional statements | formalized semantic distinction between assertional and universal statements; contingent statements are not modelled in OWL | no formalized semantic distinction |

One of the major differences between an RDF-based graph and a labeled property graph is that OWL can be mapped to RDF and thus provides formal semantics and reasoning following a W3C recommendation. No W3C-recommended standardized mapping exists for property graphs such as Neo4j so far, although proposals have been made (e.g., OWLStar, owl2lpg). OWL is based on Description Logics, and thus strictly separates **universal statements**[1] in the form of **class-specifications** (terminology box; **TBox**) from the **domain of discourse** in the form of **assertional statements**[2] that relate instances (i.e., individuals) to each other (assertion box; **ABox**) (38). Because, in OWL, classes possess URIs and can thus be referenced, but none of their class axioms do, TBoxes do not belong to the domain of discourse—within a knowledge graph, it is not possible to refer to a particular axiom of a given class but only to the class itself. Therefore, it is not possible to make statements about class definitions (i.e., universal statements) in RDF/OWL-based knowledge graphs. Consequently, documenting statements such as '*author A asserts universal law X*' is not straightforward in a knowledge graph if the universal law *X* is modelled as a class.

---

[1] A universal statement is a statement that is true for every instance of a specific universal. Universal statements represent commonly accepted domain knowledge and specify what is necessarily the case. Universal statements are based on *all-to-some* relations. Definitions of ontology classes and their class axioms are examples of universal statements, but also many scientific hypotheses and law-like regularities take the form of universal statements.

[2] An assertional statement is a statement that is true for specific particulars and specifies a fact. Assertional statements are based on *one-to-one* relations. In a knowledge graph, assertional statements relate particular instances to each other. Empirical data are examples of assertional statements.

On the other hand, because property graphs typically do not distinguish between ABox and TBox expressions, it is not always clear what their statements actually mean. For instance, the statement '*hand has part thumb*' (Fig. 5A) could refer to a particular hand and its thumb, e.g., the right hand of one of the authors of this paper. As such, the statement would represent an assertional statement. However, the statement could also refer to some unknown hand in the sense that it states that there exists at least one hand that has some thumb as its part, which would represent a **contingent statement**[3]. Another possible interpretation of the statement could be that it documents a situation typically found when dealing with hands: a hand typically possesses a thumb. In this case, the statement would represent a **prototypical statement**[4], which is a special case of a contingent statement. Finally, the statement could refer to all instances of a class hand (FMA:9712) and represent a universal statement, indicating that every hand necessarily possesses a thumb. In other words, due to the lack of formal semantics, labeled property graphs provide no standardized way to distinguish between universal, contingent, prototypical, and assertional statements.

Regarding universal statements, from a logical point of view, it makes no sense to relate two classes such as hand (FMA:9712) and thumb (FMA:24938) to each other via the property *hasPart* (BFO:0000051), because classes cannot have other classes as their parts. With few exceptions (e.g., *subclassOf* (RDFS:subclassOf), *type* (RDF:type)), object properties model instance-instance and not class-class or instance-class relations. If such object properties are used nevertheless to describe relationships between classes, this leads to semantic inconsistencies that may not cause issues for human readers, but for machines such as reasoners they do.

In OWL, the universal statement '*every instance of a hand has some instance of a thumb as its part*' can be expressed in a semantically proper way using OWL-specific properties, which results in a complex but semantically precise representation of the statement (Fig. 5B). A human reader, however, typically does not want to look at this complex representation but prefers the simpler and still intuitively comprehensible representation provided by a property graph (Fig. 5A). Moreover, the OWL-based representation involves blank nodes, which makes basic SPARQL querying difficult (38) and has the potential to cause other issues (39).

---

[3] A contingent statement is a statement that is true for some instances of a specific universal and specifies what is possibly the case. Contingent statements are based on *some-to-some* relations. In a knowledge graph, contingent statements are true for some, but not necessarily every instance of a class.

[4] A prototypical statement is a statement that is true for all instances of a specific universal as long as not the contrary is explicitly stated, and thus specifies what is typically but not necessarily the case.

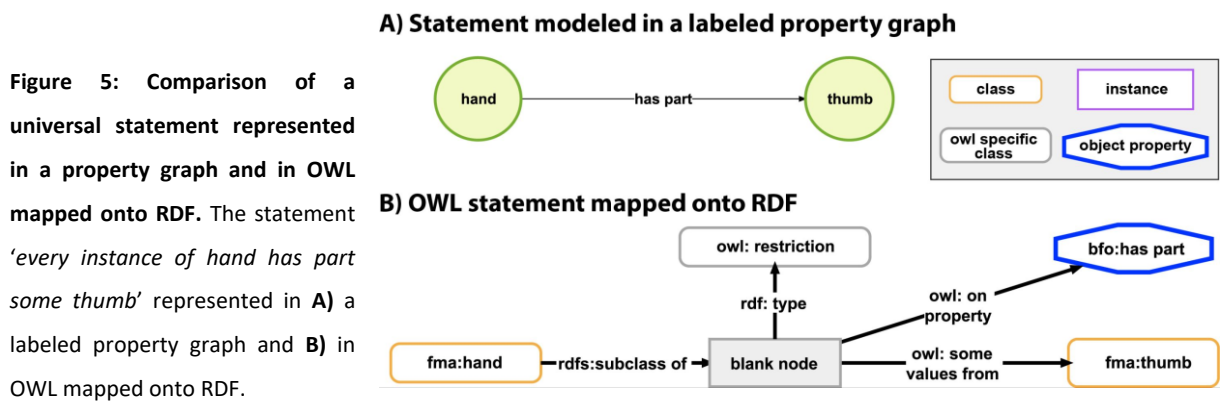

**Figure 5: Comparison of a universal statement represented in a property graph and in OWL mapped onto RDF.** The statement '*every instance of hand has part some thumb*' represented in **A)** a labeled property graph and **B)** in OWL mapped onto RDF.

Solutions for representing universal statements in a property graph have been suggested. They involve annotating edges with logical properties such as existential restriction axioms, resulting in an object-property mapping (37,38). A standardized W3C recommendation for a formalism for mapping OWL into a property graph, however, is still lacking.

Organizing a knowledge graph into different semantic units, each with its own UPRI and classified as either universal, contingent, or assertional statement, would not only provide a formal conceptual framework that facilitates the application of such object-property mappings, but would allow making statements about universal statements, which would make universal statements accessible to the domain of discourse, meaning formal class definitions (i.e., class axioms) would become referenceable so that one could make statements about them within the graph.

# Results

## Organizing knowledge graphs into semantically meaningful units of representation

To make knowledge graphs more manageable and provide them with increased levels of structure, their graph must be organized into several layers of partially overlapping, partially enclosed, subgraphs. Approaches for systematically structuring a knowledge graph into subgraphs have been proposed before, e.g., defining a **characteristic set** as a subgraph consisting of all triples that share the same resource in the *Subject* position, yielding significant improvements in space and query performance (40,41) (see also the related concept of **RDF molecules** (42,43)).

However, instead of structuring a knowledge graph into subgraphs based on the rather technical aspects of shared graph-topological properties of its triples, we here propose an approach that focuses on **structuring a knowledge graph into identifiable and semantically meaningful sets of triples, i.e., subgraphs that represent units of representation (short: semantic units)**. The approach builds on an idea suggested in the context of semantic data models for structuring descriptions of phenotypes into smallest semantically meaningful units, i.e., assertions, each of which is organized in its own Named Graph (44). Here, we generalize and extend this idea to the concept of semantic units that can be accessed, searched, and reused as identifiable and reusable data items in their own right, forming units of representation that are FAIR Digital Objects that can be implemented using RDF/OWL-based graphs as well as property graphs.

A semantic unit is a subgraph of the overall knowledge graph that represents **a unit of information that is semantically meaningful to a human reader**. Every semantic unit is represented in the graph by its own resource (i.e., node) and thus possesses its own UPRI. Moreover, it can be classified as an instance of a respective ontology class.

In a knowledge graph that is structured into semantic units, two different graph layers are distinguished: a data graph layer and a semantic-units graph layer. The **data graph layer** contains all the triples of a knowledge graph that semantic units organize into identifiable subgraphs. A knowledge graph that is not structured into semantic units only possesses a data graph layer. The **semantic-units graph layer** is added to a knowledge graph with organizing it into semantic units and comprises all the triples added to the graph due to organizing and structuring it into semantic units. The resource of each semantic unit and their interrelationships are for instance part of the semantic-units graph layer. Analogously to this general distinction across the knowledge graph, we can distinguish a data graph and a semantic-units graph for each semantic unit. The data graph of a particular semantic unit possesses the same UPRI as its semantic unit resource, so that by referring to the UPRI one refers at the same time to the semantic unit as a resource as well as its corresponding data graph and thus can make statements about the data graph's content (Fig. 7).

Two basic categories of semantic units and their subcategories can be distinguished: **statement units** and **compound units** (for a classification of semantic units, see Fig. 6).

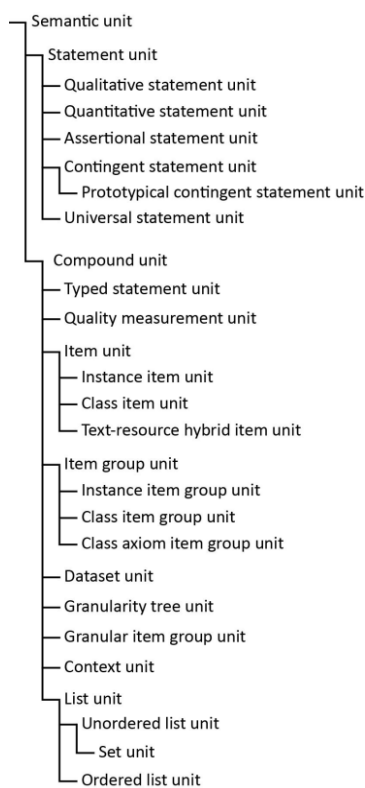

Figure 6: Classification of different categories of semantic unit.

## Statement unit

A statement unit is characterized as a **unit of information that represents the smallest, independent proposition (i.e., statement) that is semantically meaningful for a human reader** (see also (44)). For example, the fact that a population infected with a virus has the quality basic reproduction number (NCIT:C173777) represents a statement unit that is independent of an actual measurement of the basic reproduction number, which represents another statement unit. The measurement statement unit is independent of the quality statement unit because a given population may have more than one measurement for its basic reproduction number for the same point in time, due to measurement errors, measurement inaccuracies, or the application of different measurement methods, instruments, or models. The same applies to all quality measurements. For instance, the weight measurement for *ObjectX* shown in Figure 2 consists of two statement units. One statement unit specifies that *ObjectX* has the quality 'weight' (PATO:0000128), whereas the other specifies the weight's measurement.

Structuring a knowledge graph into statement units results in a **mathematical partition of its data graph layer**, with each triple of the data graph layer belonging to exactly one statement unit. It is important to note that only the data graph layer is partitioned into smaller data (sub)graphs. All the triples added to the knowledge graph for organizing it into semantic units, i.e., the semantic-units graph layer, are excluded from the partition because otherwise applying semantic units would result in an infinite regression, since adding a semantic unit always adds triples to the semantic-units graph layer, which, if included in the partition, would have to be represented themselves via a statement unit, which would add triples to the semantic-units graph layer again, etc.

Following the **predicate-argument-structure** from linguistics, the main verb of a statement, together with its auxiliaries, represents the statement's predicate. Each predicate has a **valence** that determines the number and types of **arguments** it requires to complete its meaning. **Adjuncts** can additionally relate to the predicate, but they are not necessary for completing the predicate's meaning and only provide optional information, such as a time specification in a has-part statement. Subject and object phrases are the most frequently occurring arguments and adjuncts. So, we can say that each statement unit documents a particular proposition by relating a resource that is the subject argument of the predicate of the proposition to some literal or to another resource, which is the object argument or object adjunct of the predicate. The subject argument of a proposition of a statement unit we call **subject** and the object argument(s) and object adjunct(s) the **object(s)** of the statement unit. Every statement unit has one such subject and one or more objects.

A has-part statement unit (Fig. 7), for instance, has a subject and one object as an argument, whereas a weight measurement statement unit has a subject and two objects as arguments, i.e., the weight value and the weight unit. The resource representing a statement unit in the graph relates to its subject via the property *hasSemanticUnitSubject*, which is documented in the semantic-units graph.

If the proposition documented in the data graph is based on a **binary relation**, and thus a divalent predicate, such as '*Lars' right hand has part Lars' right thumb*', the corresponding statement unit typically consists of a single triple, because in RDF, the *properties* of triples are binary relations. However, many propositions are based on **n-ary relations** and thus cannot be modelled with a single triple. For instance, the statements '*a weight measurement with the value 5 and the unit kilograms*' and '*Lars' right hand has part Lars' right thumb on January 29$^{th}$ 2022*' consist of at least two triples, whereas the statements '*Anna gives Bob a book*', '*Carla travels by train from Paris to Berlin on the 29th of June 2022*', and '*Dave buys the car from Eric*' each require more than two triples. Depending on the relation of the underlying proposition and thus the valence of its predicate and depending on the number of optional adjuncts that should be additionally modelled, statement units can thus consist of one or more triples. A specific type of statement unit, i.e., a statement unit class, could thus be defined in reference to the n-ary property that models its underlying relation. Many object properties of the Basic Formal Ontology 2 are actually ternary relations, as they are time-dependent (45). For instance, '*b* located_in *c* at *t*' requires at least two triples to be modelled in RDF.

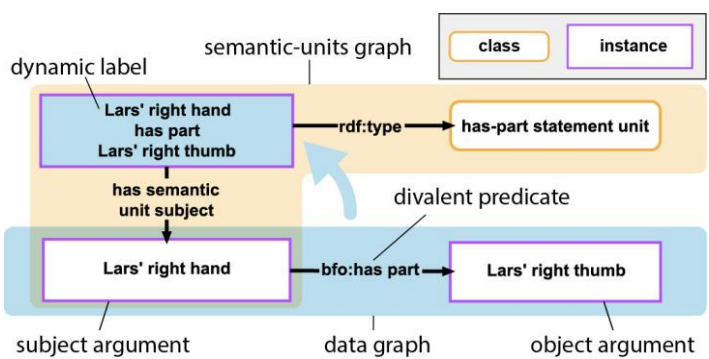

**Figure 7: Example of a statement unit.** Example of a semantic unit that models a statement based on a has-part-of relation between two instances. The data graph expressing the statement is shown in the blue box at the bottom, with 'Lars' right hand' being the subject argument and 'Lars' right thumb' the object argument of the statement. The semantic-units graph is shown in the peach-colored box and contains the triples representing the semantic unit. It specifies that the resource representing the statement unit (blue box with borders, here shown with its dynamic label), which represents the statement in the data graph (indicated by the blue arrow), is an instance of *has-part-of statement unit* and that 'Lars' right hand' is the subject of this semantic unit. The UPRI of *'has-part-of statement unit'* is also the UPRI of the semantic unit's data graph (the subgraph in the blue box without border).

In other words, which triples belong to a statement unit must be specified for each type of underlying predicate on a case-by-case basis by human domain experts. However, by specifying a graph pattern for modelling a given proposition into a graph-based representation that in turn constitutes a statement unit, modelling languages such as LinkML or the Shapes Constraint Language SHACL (32) can be used, and corresponding tools can be applied for identifying instances of this pattern and thus instances of the corresponding type of statement unit. Moreover, the SHACL shapes can be utilized to derive **dynamic labels** for each statement unit when representing them in visualizations of the knowledge graph in a user interface (UI). Such a label can be generated dynamically by parsing the label from the statement unit's subject resource and the labels from all of its object resources and literals and creating from them a human-readable statement following a pre-defined template. E.g., the template '*subject* travels by *A* from *B* to *C* on the *D*', with the input '*Carla | train | Paris | Berlin | 29th of June*' would read as '*Carla* travels by *train* from *Paris* to *Berlin* on the *29th of June 2022*'.

Each statement unit instantiates a respective statement unit ontology class. Characteristic for the class is the type of relation of the propositions its instances are modelling. Based on the relation of its underlying proposition, one can distinguish two subcategories of statement unit (**relation-based criterion**):

1. **Qualitative statement unit**: Statement units that document qualitative relations, such as a has-part relation (Fig. 7). The **argument objects of a qualitative statement unit are always**

> **resources and not numerical values**—only adjunct objects of a qualitative statement unit may be numerical values.

2. **Quantitative statement unit**: Statement units that document quantitative statements, such as a measurement datum. **At least one argument object of a quantitative statement unit must be a numerical value**.

The two categories are exhaustive and disjoint, meaning that any given statement unit is either a qualitative or a quantitative statement unit, but never both. Within these two subcategories, based on the underlying relation, e.g., *has part*, *type*, *develops from*, one can differentiate between further statement unit classes. Additional subtypes of a given relation-based statement unit class, e.g., has-part statements about material entities versus has-part statements about information content entities, do not have to be differentiated as classes, but can be identified and distinguished by simply querying for the *\*hasSemanticUnitSubject\** relations that instances of a has-part statement unit class have to their subject.

Besides this relation-based criterion, there is another criterion for differentiating classes of statement units. The **subject-category-based criterion** is based on the category of entity of the statement unit's subject, resulting in the distinction of assertional, contingent, and universal statement units. These three categories of statement units are exhaustive and disjoint, meaning that any given statement unit is either an assertional, a contingent, or a universal statement unit, but never a combination of them (see also *Challenge 4*). Since the relation-based and the subject-category-based criterion are mutually independent, a given statement unit can, for instance, be a weight measurement statement unit and an assertional statement unit.

Before we discuss the different types of subject-category-based statement units, we first want to point to a convention we want to introduce, which we apply to semantic units to contribute to closing the conceptual gap discussed in *challenge 4*. This convention involves the introduction of two new types of resources in addition to instances, properties, and classes: **some-instance and every-instance resources**. Based on these two new types of resources, three different types of qualitative statement units can be distinguished, which model three different types of relations between some-instance, every-instance, named-individual, and class resources in their data graph.

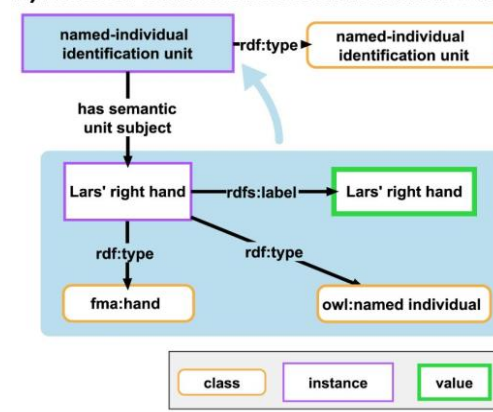

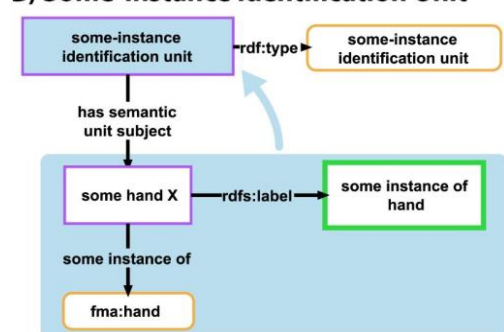

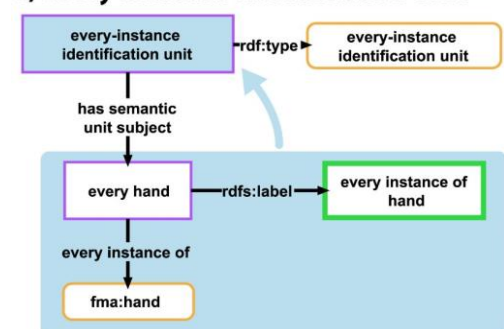

**Figure 8: Examples for three different types of identification units. A) Named-individual identification unit**. This named-individual identification unit models in its data graph the class-affiliation of the instance 'Lars' right hand' (FMA:9712), which is its subject, through the property *type* (RDF:type), together with the subject's label 'Lars' right hand' through the property *label* (RDFS:label). The blue box without borders shows the data graph that belongs to this named-individual identification unit (blue box with borders). Subjects of named-individual identification units are the type of instances that are also the subjects of assertional statement units (cf. Fig. 9A). **B) Some-instance identification unit**. This some-instance identification unit models in its data graph the class-affiliation of the instance 'some hand X' (FMA:9712), which is its subject, through the property \**some instance of*\*, together with the subject's label 'some instance of hand'. The blue box without borders shows the data graph that belongs to this some-instance identification unit. Subjects of some-instance identification units are the type of instances that are also the subjects of contingent statement units (cf. Fig. 9B). **C) Every-instance identification unit**. This every-instance identification unit models in its data graph the class-affiliation of the instance 'every hand' (FMA:9712), which is its subject, through the property \**every instance of*\*, together with the subject's label 'every instance of hand' through the property *label* (RDFS:label). The blue box without borders shows the data graph that belongs to this every-instance identification unit. Subjects of every-instance identification units are the type of instances that are also the subjects of universal statement units (cf. Fig. 9C).

1. **Named-individual identification unit**: A qualitative statement unit that identifies a resource that is the subject of the statement unit to refer to a named individual, together with its label and its type, i.e., its class membership (Fig. 8A).

2. **Some-instance identification unit**: A qualitative statement unit that identifies a resource that is the subject of the statement unit to refer to some unspecified instances of a given class, together with a label and its class specification (Fig. 8B). When this resource is used in some other statement unit, it is meant as '∃i∈C' (with i = instance and C = class), reading 'there exists an instance i of class C for which holds ...'. A some-instance identification unit may, additionally, also contain information about cardinality restrictions (see [Modelling negations, cardinality restrictions, and disagreement](#) below).

3. **Every-instance identification unit**: A qualitative statement unit that identifies a resource that is the subject of the statement unit to refer to every instance of a given class, together with a label and its class specification (Fig. 8C). When this resource is used in some other statement unit, it is meant as '∀i∈C' (with i = instance and C = class), reading 'for every instance i of class C holds'.

**Assertional statement unit**

An assertional statement unit is a statement unit that has a **named individual** as its subject-resource (e.g., 'Lars' right hand' (FMA:9712) in Fig. 8A), and thus an instance that is known to the knowledge graph. In other words, its subject is also the subject of a named-individual identification unit. If the assertional statement unit possesses an object that is a resource and not a literal, that object also refers to a named individual, resulting in a *one-to-one* relation (or many such relations in case of an n-ary proposition). An assertional statement unit is thus an ABox representation of a relation with at least one particular.

As a semantic unit, a particular assertional statement unit refers to a subgraph of a knowledge graph that models a particular assertion. An assertion is a proposition about a particular (the subject) and thus about an individual and not a universal, and this proposition is meant to be either *true* or *false* (46,47). In a knowledge graph, a particular assertional statement unit is modelled by a resource that represents the unit's data graph. This data graph is thus an instance-based subgraph of the knowledge graph and represents an **ABox expression**. The unit resource is an instance of both the respective type of statement unit and of *assertional statement unit* (Fig. 9A).

**Contingent statement unit**

A contingent statement unit is a statement unit that refers to **some instances of a specific class** as its subject. In other words, its subject is also the subject of a some-instance identification unit. If the contingent statement unit possesses an object that is a resource, that object also refers to some instances of a specific class, resulting in a *some-to-some* relation (or many such relations in case of an n-ary proposition). For example, if you want to state that some but not necessarily all hands possess a thumb, and you do not want to or simply cannot list all the individual hands to which this applies, you make such a contingent statement unit (Fig. 9B).

Contingent statements are propositions that are true for some instances of a specific universal, but not necessarily for every instance (46–48). They are frequently used when expressing **prototypical relationships**, such as that a human's hand usually has a thumb as its part, and thus statements about universals that are assumed to be at least typically true in specific contexts, but not necessarily so—a hand is still a hand, even if it has no thumb[5]. Other examples are statements about the relationships between diseases and their symptoms or drugs and their effects, which are often expressed in reference to some probability evaluation. These prototypical contingent statements can be documented in **prototypical contingent statement units**, which form a subcategory of contingent statement units.

Description Logics and thus also **OWL does not provide the means to state what is typically *true***, and other logical formalisms that are capable of expressing prototypical knowledge are needed to provide a semantics for such statements (47).

In a knowledge graph, a particular contingent statement unit can be modelled as a statement unit, with its resource representing the corresponding data graph. The resource is an instance of both the respective type of statement unit and of *contingent statement unit*. Its subject and, if present, also its object is modelled as a resource that relates to its class through a *some instance of* property in corresponding some-instance identification units (Fig. 8B). Probability values expressing an assumed value of the frequency of instances for which the statement is *true* can be documented using additional triples, forming in turn an assertional statement unit in its own right, which would have the contingent statement unit's resource as its subject.

---

[5] Therefore, the example given in Fig. 9C is a statement that human anatomists would not make. They would define a human hand in reference to other properties.

**Figure 9: Three subcategories of statement unit. A) Assertional statement unit.** An example of a statement unit that is an *assertional statement unit*. It models a has-part relation between two instances. The data graph expressing this relation is shown in the blue box without borders. It is an instance-based data graph and forms an ABox. The subject of this assertional statement unit is an instance resource that is also the subject of a named-individual identification unit (cf. Fig. 8A). The resource representing this data graph, i.e., the assertional statement unit resource (blue box with borders, here shown with its dynamic label), is an instance of *has-part statement unit* and of *assertional statement unit*, which is represented in the semantic-units graph above the blue box without borders. **B) Contingent statement unit.** An example of a statement unit that is a *contingent statement unit*. It models a has-part relation that exists between some but not necessarily every instance of 'hand' (FMA:9712) and some but not necessarily every instance of 'thumb' (FMA:24938). The data graph expressing this relation is shown in the blue box without borders. The subject of this contingent statement unit is an instance resource that is also the subject of a some-instance identification unit (cf. Fig. 8B). The resource representing this data graph, i.e., the contingent statement unit resource (blue box with borders, here shown with its dynamic label), is an instance of *has-part statement unit* and of *contingent statement unit*. **C) Universal statement unit.** An

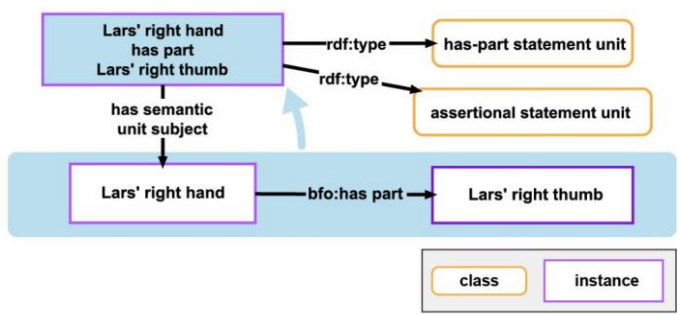
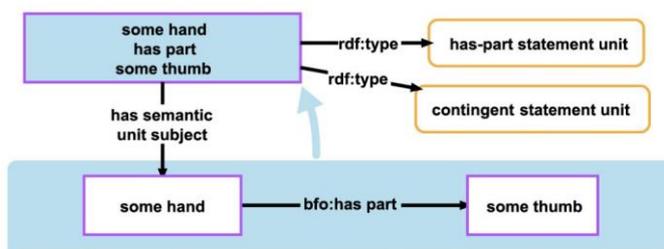
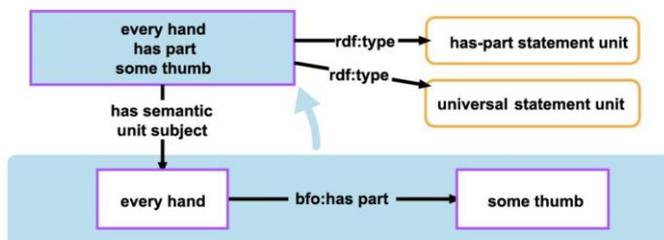
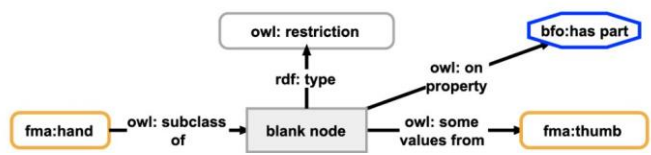

example of a statement unit that is a *universal statement unit*. It models a has-part relation that exists between every instance of 'hand' (FMA:9712) and some, but not necessarily every instance of 'thumb' (FMA:24938). The data graph expressing this relation is shown in the blue box without borders and can, in principle, be translated into OWL to form a TBox. The subject of this universal statement unit is an instance resource that is also the subject of an every-instance identification unit (cf. Fig. 8C). The resource representing this data graph, i.e., the universal statement unit resource (blue box with borders, here shown with its dynamic label), is an instance of *has-part statement unit* and of *universal statement unit*. **D) Conventional OWL model.** The example from C) translated into an OWL class-axiom expression mapped to RDF.

**Universal statement unit**

A universal statement unit is a statement unit that refers to **every instance of a specific class** as its subject. In other words, its subject is also the subject of an every-instance identification unit. If the universal statement unit possesses an object that is a resource, the object refers to some instances of a specific class, resulting in an *all-to-some* relation (or many such relations in case of an n-ary proposition). If you want to state that every instance of hand (FMA:9712) necessarily *hasPart* (BFO:0000051) some thumb (FMA:24938), you make a universal statement (Fig. 9C). Universal statements represent commonly accepted domain knowledge that take the form of, for instance, definitions of terms, expressed as class axioms, or of hypotheses about law-like causal relationships.

As a semantic unit, a particular universal statement unit refers to a data graph within a knowledge graph that represents a particular universal statement and thus a proposition that is *true* for all instances of a specific universal (46,47). In a knowledge graph, a particular universal statement unit is modelled as a statement unit by a resource that represents the corresponding data graph. The resource is an instance of both the respective type of statement unit and of *universal statement unit* (Fig. 9C). Its subject is modelled as a resource that relates to its class through an *every instance of* property in a corresponding every-instance identification unit (Fig. 8C), whereas its resource objects, if present, are modelled as resources that relate to their class through the *some instance of* property in a corresponding some-instance identification unit (Fig. 8B).

When compared to the OWL modelling of universal statements, which is rooted in Description Logics (cf. with Fig. 9D), the data graph belonging to a universal statement unit as a semantic unit is less complex and does not include any blank nodes. This improves its querying properties. Moreover, the statement unit's resource can be used to make statements about a universal statement, making universal statements and thus also particular class axioms available to the general domain of discourse (see *Challenge 4*). However, because no reasoner yet exists for directly reasoning with this semantic-units-based notation, to be able to do reasoning over these statements, they will have to be translated into an OWL-based notation, forming a classical **TBox expression** (see Logical semantics of semantic units).

## Compound unit

Compound units are semantic units that organize the set of all statement units of a given knowledge graph into larger but still semantically meaningful data graphs. Technically, a compound unit is a container of resources of particular semantic units and thus specifies a collection of associated semantic units. Merging the data graphs of its associated semantic units constitutes the data graph of the compound unit. Analog to statement units, each compound unit is a semantic unit that is represented in the graph by its own node, possesses its own UPRI, and instantiates a corresponding compound unit ontology class.

In the semantic-units graph of a compound unit, the relation between the resource representing the compound unit and the resources representing its associated semantic units is documented via the property *hasAssociatedSemanticUnit* (Fig. 10). The following subcategories of compound unit can be distinguished.

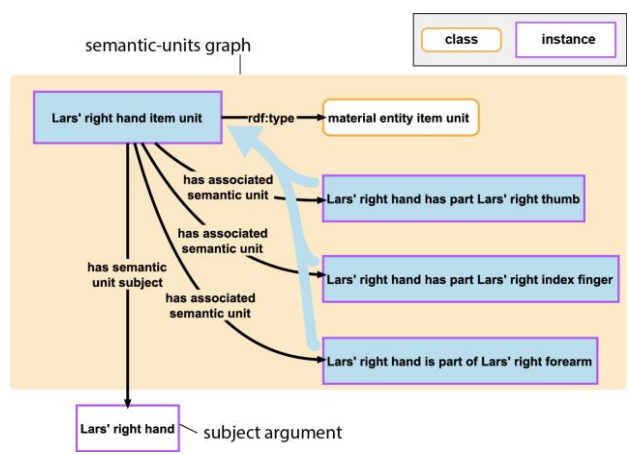

**Figure 10: Example of a compound unit** that comprises several statement units. Compound units possess only indirectly a data graph, through merging the data graphs of their associated statement units. The compound unit resource (here, *'Lars' right hand item unit'*), however, stands for these merged data graphs (indicated by the blue arrow). Compound units possess a semantic-units graph (shown in the peach-colored box), which documents the semantic units that are associated with it.

**Typed statement unit**

In order to be readable for a human being, the URIs of the subject, predicate, and objects of a statement unit must be translated to their respective labels. The dynamic label of a statement unit does exactly that, in a structured way, representing the statement itself in a human-readable plain English sentence. The information contained in a dynamic label, however, goes beyond the information contained in the statement itself, because it includes information from the identification units associated with the statement's subject and its object resources. Since identification units can be edited independently of the statement units that use their resources as subjects and objects, when referring to a human-readable statement, one is actually referring to the corresponding statement unit and the identification units of its subject and all of its objects. This collection of semantic units is a typed statement unit. A typed statement unit is a compound unit that comprises the following statement units (Fig. 11):

1. A statement unit that is not an instance of one of the three types of identification units (see Fig. 8). This statement unit functions then as the **reference statement unit** of the typed statement unit, and its subject is also the subject of the typed statement unit to which it relates via the property *hasSemanticUnitSubject*.

2. The **identification units** that model the class-affiliations of all the resources that are referenced in the data graph of the reference statement unit.

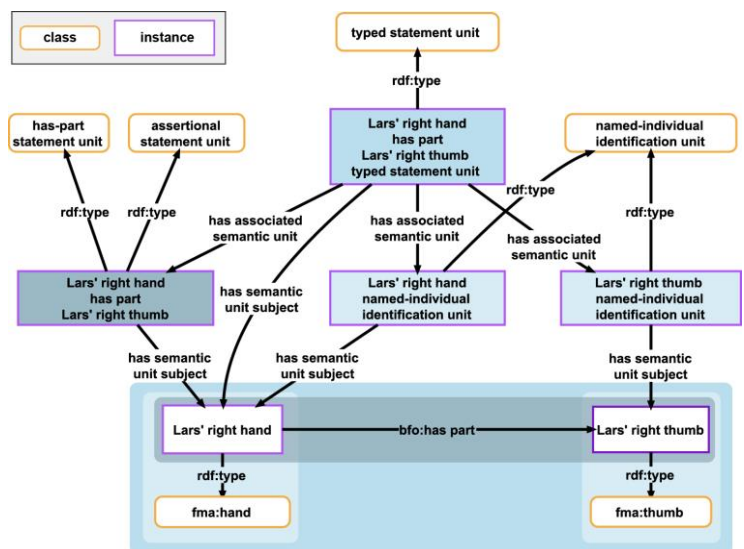

**Figure 11: Typed statement unit.** An example of a typed statement unit, with the data graph shown in the large blue box without borders. This graph results from merging the subgraphs of three statement units: the *'part-of assertional statement unit'* from Fig. 9A, which is the reference statement unit of this *'typed part-of statement unit'*, with its data graph shown in the grey box without borders, and two instances of *named-individual identification unit*, with their data graphs in the light-blue box without borders (the resources representing the different data graphs are shown in correspondingly colored boxes with borders and dynamic labels).

Typed statement units are important because they contain class-affiliation and label information that human readers usually require for understanding the proposition of the reference statement unit. Since the resource representing a typed statement unit is related to its reference statement unit and all its identification units through the property *hasAssociatedSemanticUnit*, different types of typed statement units can be identified by simply querying these relationships, eliminating the need to further differentiate the class of typed statement units into subclasses.

**Quality measurement unit**

A quality measurement unit has one qualitative typed statement unit and one or more quantitative typed statement units associated via the property *hasAssociatedSemanticUnit*. The reference statement unit of its qualitative typed statement unit specifies a quality of a given entity, whereas the statement unit of each of its quantitative typed statement units specifies a measurement of this quality, comprising a value and a unit. A quality measurement unit is thus a collection of typed statement units that describe a specific quality of a given entity together with all quantitative evaluations or measurements of it currently present in the data graph (the graph shown in Figure 2 corresponds with a typical data graph of a quality measurement unit).

For example, the basic reproduction number of a given population at a given time can be documented in such a quality measurement unit, which would comprise the typed qualitative statement unit that links the population with the quality basic reproduction number (NCIT:C173777) and one or more typed quantitative statement units, each of which documents a measurement of this reproduction number, maybe based on different methods and algorithms for its estimation.

The basic reproduction number measurement statement unit takes the object resource of the quality statement unit and links the quality statement unit to the basic reproduction number measurement statement unit via the *objectDescribedBySemanticUnit* property. The quality statement unit in its turn has taken the subject of the quality measurement unit it is associated with as its subject and is linked to the respective quality measurement unit via the property *hasAssociatedSemanticUnit*, as is the basic reproduction number measurement statement unit.

The subject of the quality statement unit is also the subject of the quality measurement unit, and the resource representing a quality measurement unit in the semantic-units graph relates to this resource via the property *hasSemanticUnitSubject*.

Because the resource representing a quality measurement unit in the semantic-units graph is related to its associated typed statement units and their reference statement units through the property *hasAssociatedSemanticUnit*, different types of quality measurement units can be identified by simply querying these relationships, eliminating the need to further differentiate the class of quality measurement units into subclasses.

**Item unit**

An item unit comprises all statement units, typed statement units, and all quality measurement units that share the same subject. The corresponding resource is then also the **subject of the item unit**, and the resource representing an item unit in the semantic-units graph relates to its subject via the property *hasSemanticUnitSubject*. An item unit is thus a collection of semantically related statement units, typed statement units, and quality measurement units that all contain information about the same entity. Conceptually, item units relate to the *graph-per-resource* data management pattern (49) or to the abovementioned *characteristic set* or *RDF molecule*, but applying its idea to statement units instead of triples, similar to *Item* in the Wikibase data model.

The information contained in an item unit is well suited to be presented on the same page in a UI, as the statement units, typed statement units, and quality measurement units belonging to it describe the same entity, i.e., their shared subject. The following subcategories of item unit can be distinguished:

1. **Instance item unit**: An item unit that has a **named individual** resource as its subject. Consequently, it consists exclusively of all assertional statement units present in the semantic-units graph layer of the knowledge graph that share the same subject. Its subject is also necessarily the subject of a named-individual identification unit. For example, all statements about a particular infected population, such as the population's location, its qualities, various measurements about it, including basic reproduction number and case fatality rate, as well as the time period to which the information about the population refers, all are typed statement units and quality measurement units that share the same subject, which is a named individual. Together, they therefore constitute an instance item unit. The information contained in an instance item unit can be presented in the same UI page. Another example is the description of the anatomy of a particular specimen, where one could organize all typed statement units and quality measurement units describing a particular part of the specimen in an instance item unit, resulting in several such instance item units—one for each described part of the specimen.

2. **Class item unit**: An item unit that has a **some-instance** or an **every-instance resource of a specific class** as its subject. Consequently, it consists exclusively of all universal and all contingent statement units present in the semantic-units graph layer of a knowledge graph that share the same subject. Simple star-shaped class axioms such as the one shown in Figure 12 can be organized in such a class item unit.

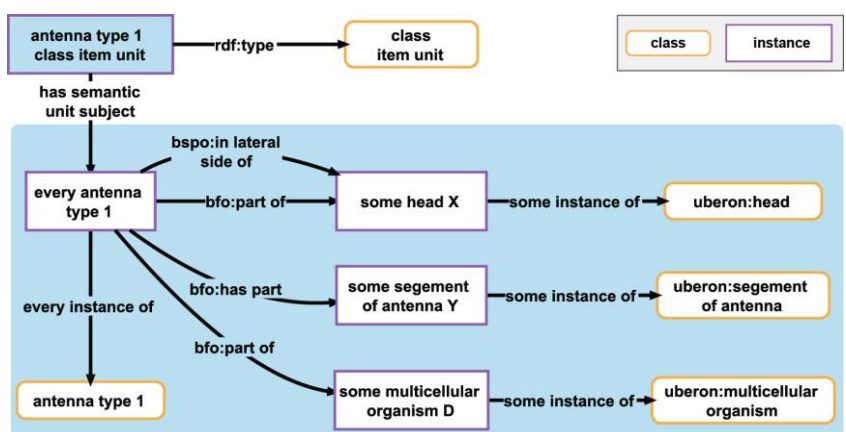

**Figure 12: Class item unit.** An example of a class item unit (blue box with borders), with its data graph shown in the blue box without borders. The data graph results from merging the data graphs of four universal statement units that have *'every antenna type 1'* as their semantic unit subject, together with their corresponding identification units.

*For reasons of clarity of presentation, resources and relations of associated semantic units not shown.*

3. **Text-resource hybrid item unit**: Text-resource hybrid item units are used when a resource needs to be described and one does not want to or cannot describe it properly in a formalized machine-actionable semantic way using triple statements and proper terms from ontologies. For instance, in a crowdsourced knowledge graph, when a user wants to describe a specific research objective or the main contribution of a research paper. This user is a domain expert, but most likely with only limited knowledge about semantics, if at all. In such a case, it might be best to specify the research objective or main contribution as a resource that is an instance or subclass of, e.g., <u>objective specification</u> (IAO:0000005), and provide a textual description for it, instead of describing it formally using triples. The textual description can either be documented as the resource's description through a respective annotation property with a string as its value or, in Neo4j, as a value node that <u>denotes</u> (IAO:0000219) the resource. In a knowledge graph application, this textual description can then be automatically annotated semantically by recognizing mentioned entities that are *known* to the knowledge graph. These recognized entity resources can be connected to the '<u>objective specification</u>' resource via the relation <u>mentions</u> (IAO:0000142). The resulting set of triples forms a data graph representing a text-resource hybrid item unit. Text-resource hybrid item units provide human-readable information in a textual description that can also be leveraged to some extent by machines via the <u>mentions</u> relations.

**Item group unit**

An item group unit consists of at least two item units (or an item unit and a typed statement unit that does not share the same subject) that are semantically related to one another through statement units that relate their subjects in the following way:

    IF    *resourceX* = subject of *itemUnit1*,

            *resourceX* = subject of *statementUnit1*,

            *resourceY* = object of *statementUnit1*,

            *resourceY* = subject of *itemUnit2*

    THEN    *statementUnit1* semantically connects *itemUnit1* and *itemUnit2* to constitute an item group unit.

We refer to statement units that link two item units in the way here described as **item links**. In the semantic-units graph layer, this is indicated by relating the resource of *itemUnit1* to that of *itemUnit2* via the property *<u>hasLinkedSemanticUnit</u>* and the linking statement unit to *itemUnit2* via the property *<u>objectDescribedBySemanticUnit</u>*. While each individual item link has a specific direction in which it links two item units, two item units can be linked through more than one item link, the directions of which can be opposite. For example, when describing a family with all its members, the item unit describing the father links to that of the daughter as his daughter, whereas the item unit describing the daughter links to that of the father as her father. Consequently, there is no generally applicable rule for identifying a specific resource as the subject of an item group unit, and thus no subject is specified.

In addition to item units, item group units may also consist of individual statement units, typed statement units, and quality measurement units that are not part of any specific item unit, because no other statement unit shares the same subject.

In most of the cases, item group units will contain all data about the same object, with some of its item units containing data related to the parts or specific aspects of this object. For example, all data about a particular scholarly publication can be understood to represent the outcome of a research activity. All data relating to that research activity can be organized and represented by a single item group unit, with its item units containing statement units that describe specific parts of that activity, such as smaller research steps, methods that have been applied, materials, instruments, and research agents that have been involved in that research activity.

With the information from each item unit being presented in the UI in its own page, item group units can be accessed by a human reader in the UI as a set of linked pages.

One can differentiate item group units into instance item group units, class item group units, and class axiom item group units. An **instance item group unit** has only semantic units associated, whose subjects refer to **named individuals**. A **class item group unit**, on the other hand, has only semantic units associated whose subjects refer to **some-instance** or **every-instance resources of a specific class**. The semantic units associated with a class item group unit give a formal description of a specific class, which in turn is the subject of the class item group unit. If the subject of the class item group unit is also the subject of an every-instance identification unit, it is a **class axiom item group unit**. Class axiom item group units are formalized class descriptions that cannot be expressed with a single class item unit because they include chains of relations and thus comprise universal and contingent statement units that do not share the same subject, and thus must be modelled as item group units (Fig. 13).

As already discussed above, in Description Logics and thus in OWL, universal statements are modelled in a TBox which, when mapped onto RDF, includes the use of blank nodes (see, e.g., Fig. 5). The use of such blank nodes, which function like anonymous resources in triples (i.e., an instance of anonymous individual (OWL:AnonymousIndividual) in OWL2), is problematic in many contexts, mainly because every entity and property represented in a TBox as an anonymous resource cannot be identified and referenced outside the respective class axiom. This results in substantial practical consequences that significantly affect the overall usability of TBox expressions for modelling empirical data. The consequences impact the findability, accessibility, explorability, comparability, expandability, reusability, and overall machine-actionability of any expression modelled as a TBox, which is why empirical data should be preferably modelled as ABox expressions (for a detailed discussion see (30)). In the context of modelling class axioms, the blank nodes cause problems when several triples should refer to the same blank node. The Description Logics upon which OWL is based uses a form of tree-model property that restricts OWL axioms to have a tree-shaped structure, so you frequently run into difficulties when describing class axioms that involve triangular relationships when using OWL (50).

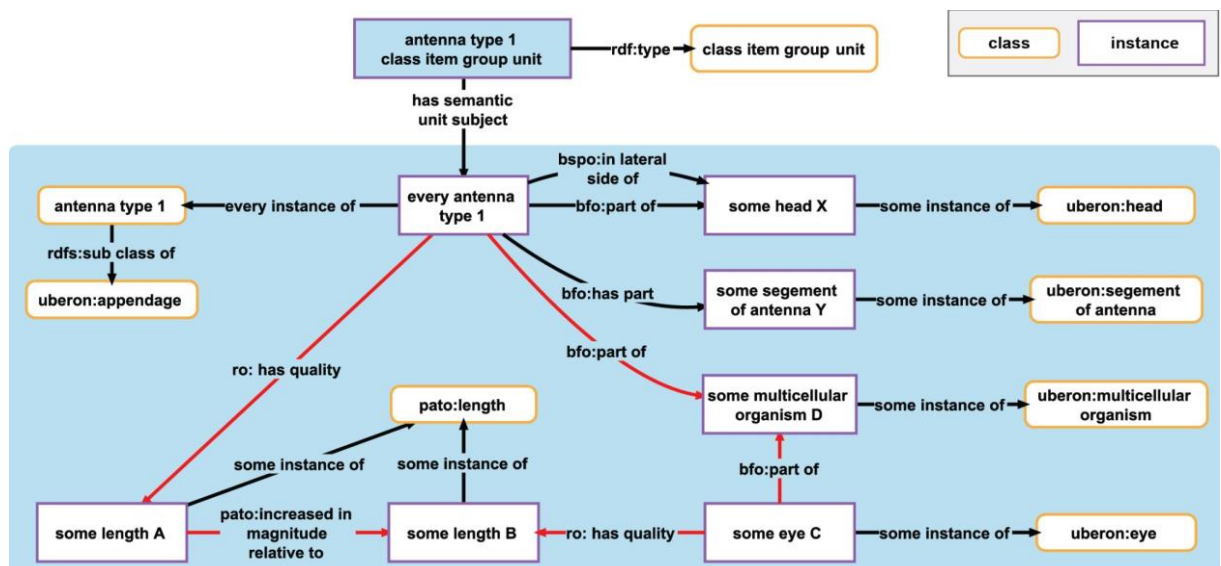

**Figure 13: Class axiom item group unit.** An example of a class axiom item group unit (blue box with borders), with its data graph shown in the blue box without borders. The data graph results from merging the data graphs of several universal statement units and their corresponding identification units. The statements cannot be modelled in a single class item unit, because they do not all share the same subject. Contrary to OWL, this notation allows describing triangular relations as the one shown here in red, indicating that every instance of *antenna type 1* is longer than any eye that is part of the same organism. *For reasons of clarity of presentation, resources and relations of associated semantic units not shown.*

For instance, if you want to formally describe a specific type of antenna that is longer than the eyes of its organism, you can try to model a respective *antenna* class in Manchester Syntax (51) with the following expression:

> 'has part some ((antenna part of **some multicellular organism**) and has quality some (length and increased in magnitude relative to some (length and inheres in some (eye part of **some multicellular organism**))))'.

However, because the two mentions of 'some multicellular organism' are anonymous, the expression does not capture the intended meaning that this should refer to the same multicellular organism, as according to it, to be an instance of the class, an antenna needs to merely be longer than at least one eye of some multicellular organism in the world, not necessarily an eye possessed by the same organism that possesses the antenna (52). Because universal statement units do not use blank nodes, different statements can refer to the same resource 'some multicellular organism D' and thus do not run into this restriction of OWL (see Fig. 13, triangular relationship shown in red).

**Granularity tree unit**

Since each statement unit is represented in a knowledge graph by its own resource (i.e., node) and instantiates a corresponding statement unit ontology class that is characterized by the type of relation that underlies the propositions its instances are modelling, we can add further information to that class. We can, for instance, identify types of statement units that depend on partial order relations. A **partial order relation** is a binary relation *p* that is transitive (if *A* has relation *p* to *B* and *B* has relation *p* to *C*, then *A* has relation *p* to *C*), reflexive (*A* has relation *p* to itself), and asymmetric (if *A* has relation *p* to *B* and *B* has relation *p* to *A*, then *A* and *B* are identical). Examples of such partial order relations are the class-subclass relations that can be found in universal statements, or parthood relations found in assertional statements, but many more partial order relations exist, including sequential relations such as *before* (RO:0002083) that identify **timelines** in a knowledge graph.

Partial order relations can give rise to granular partitions that form **granularity trees** (53–55) and can be used to define **granularity perspectives** (56–58). Granularity perspectives identify specific types of semantically meaningful tree-like subgraphs within the data graph layer of a knowledge graph and can be employed for supporting the exploration of the knowledge graph in a way somewhat orthogonally to statement, item, and item group units. By identifying classes of statement units that potentially relate to granularity perspectives, because their characteristic relations are partial order relations, applications can utilize this information for automatically identifying corresponding types of granularity trees within the knowledge graph and make them accessible for the users.

Due to the nested structure of a granularity tree and its implicit directionality from root to leaves, one can specify the **subject of a granularity tree unit** to be the subject of the statement units that share their objects with the subjects but not their subject with the objects of other statement units belonging to that granularity tree unit. The resource representing a granularity tree unit in the semantic-units graph relates to its subject via the property \**hasSemanticUnitSubject*\*.

Characteristic of granularity tree units is that their statement units are always instantiating the same statement unit class (because the class is characterized in reference to the underlying partial ordering relation) and subjects and objects are always resources of the same basic type. To make them not only machine-actionable but also better comprehensible for humans, granularity tree units comprise the typed statement units that correspond with the statement units that form the granularity tree.

Because the resource representing a granularity tree unit in the semantic-units graph is related to its associated statement units through the property \**hasAssociatedSemanticUnit*\*, different granularity perspectives can be identified and distinguished by simply querying these relationships and the \**hasSemanticUnitSubject*\* relationship to the subject of the granularity tree unit, eliminating the need to further differentiate the class of granularity tree units into subclasses.

Granularity perspectives derived from specific types of statement units thus provide another means for structuring and organizing the overall data graph into semantically meaningful smaller data graphs, and thus provide another category of compound unit. Examples for such granularity perspectives are tree-like hierarchies of has-part relations, for instance between various anatomical structures in an organism, or the temporal sequence of events of a particular process, developmental dependency chains, or more general cause-effect-relation chains, and taxonomies of concepts based on class-subclass relations.

**Granular item group unit**

A granular item group unit comprises all statement units, quality measurement units, and item units whose subjects are part of the same granularity tree unit. The subject of the granularity tree unit is then also the **subject of the granular item group unit**, and the resource representing a granular item group unit in the semantic-units graph relates to its subject via the property \**hasSemanticUnitSubject*\*.

A granularity tree provides a nested hierarchical organization of related resources. If these resources are subjects of item units, the corresponding item units can be organized according to the hierarchy provided by the granularity tree, thus organizing several item units in a nested hierarchy. This additional organization and structuring of the data graph layer can be utilized by a knowledge graph application to improve the **explorability** of its overall data graph for users. Granularity tree units can thus provide additional organizational structure for item group units by organizing their item units into granular item group units. This becomes particularly important when the knowledge graph is highly connected, and its item group units are therefore very large.

Because the resource representing a granular item group unit in the semantic-units graph is related to its associated semantic units through the property \**hasAssociatedSemanticUnit*\*, different types of granular item group units can be identified by simply querying these relationships and the \**hasSemanticUnitSubject*\* relation to their respective subjects, eliminating the need to further differentiate the class of granular item group units into subclasses.

**Context unit**

A context unit comprises all semantic units for which merging of their data graphs forms a connected graph with no resource or triple being separated from the other resources or triples, i.e. any two resources in the data graph of the context unit are connected via a series of triple statements that also belong to the data graph of the context unit. The only exception to this rule are triples that have *isAbout* (IAO:0000136) as their property and which thus belong to an **is-about statement unit**. The property *isAbout* relates an information artifact to an entity that the artifact contains information about. It thus always marks a change of frame of reference from the discursive layer to the ontological layer by having a semantic unit resource as its subject and some other non-semantic unit resource from the data graph as its object (for discussion of layers, see [Statements about statements and documenting ontological, diagnostic, and discursive information in knowledge graphs using semantic units](#)). In other words, **is-about statement units relate resources from the semantic-units graph with resources from the data graph of a knowledge graph and can be used as an index for where in the graph a change of frame of reference occurs**. For example, when documenting a research activity with its in- and output and the output being a dataset containing the description of the anatomy of a multicellular organism. The graph will contain the statement *'[description item unit](#)'* *isAbout* '[multicellular organism](#)' (UBERON:0000468), which points to a change of frame of reference from the result of the research activity in the form of a description and the multicellular organism it is describing. In short, it indicates a change from the research-activity reference frame to the research-subject reference frame (see also Fig. 14 further below). Therefore, an is-about statement unit always indicates the border between the data graphs belonging to two different context units. Similar to item group units, due to the lack of a generally applicable rule for identifying a specific resource as the subject of an item group unit, no subject is specified.

**Dataset unit**

A dataset unit is a defined and ordered collection of particular semantic units. This collection can be homogeneous, comprising only semantic units of the same type, but can also be compilations of statement units and compound units of various kinds. Like any other semantic unit, each dataset unit is represented in the semantic-units graph with its own node and its own UPRI.

Dataset units can be used in various ways. For instance, for compiling all data contributed by a specific institution within a collaborative project or for documenting the state of a given object in a given time. Also, the results of a particular search query can be stored and made accessible as a dataset unit. Users of knowledge graphs can specify particular dataset units for their own use and use the unit's UPRI as an identifier to reference it.

In addition to such obvious uses of datasets as collections of particular semantic units, dataset units can also be used for managing statements about statements: If users want to point out that a particular measurement datum (assertion statement unit *A*) from one publication (item group unit *X*) supports or directly contradicts a causal theory (universal statement unit *B*) from another publication (item group unit *Y*), they could create a respective assertion statement unit *C* that relates the statement unit *A* as supporting statement unit *B* and could create a dataset consisting of these three statement units *A*, *B*, and *C*, without including *X* and *Y*. Whereas this might not be communicated as a dataset in the UI, technically it could be structured, stored, and managed as a dataset unit.

**List unit**

Sometimes, you need to make a statement about a specific list of particular resources. To do so, such a list can be modelled as a specific type of compound unit, i.e., a list unit. List units can be utilized the same way as the other types of semantic units: each list unit is a semantic unit that is represented in the semantic-units graph by its own node, possesses its own UPRI, and instantiates a corresponding list unit ontology class.

We distinguish **unordered list units** from **ordered list units**, with the latter having the resources organized in a specific order, for instance the authors of a scholarly publication. A **set unit**, on the other hand, is an unordered list unit in which every resource is listed only once (uniqueness restriction).

Technically, a list unit is a collection of membership statement units, each of which specifies one of the resources belonging to the list by connecting the UPRI of the list unit through a *child* relation to the respective resource. In case of an ordered list unit, each membership statement unit has to be indexed first (by connecting the UPRI of the membership statement unit through a data property *index* (RDF:index) to a consecutive integer value, starting with 0), resulting in an indexed membership compound unit, which is then associated with the list unit.

List units can be used as arrays and may include cardinality restrictions and thus describe a closed collection of entities, making a localized closed-world assumption.

# Discussion

## Metadata and Semantic Units as FAIR Digital Objects

The concept of a **FAIR Digital Object** has been suggested by the European Commission Expert Group on FAIR Data as central to the realization of FAIR (16). FAIR Digital Objects must be accompanied by persistent identifiers, metadata, and contextual documentation to enable their reliable discovery, citation, and reuse. Data belonging to a FAIR Digital Object should be stored in a common and ideally open file format and be richly documented using metadata standards and well-established vocabularies.

FAIR Digital Objects also represent the core concept in the interoperability framework of the European Open Science Cloud (EOSC) as atomic entities for a FAIR ecosystem by providing the metadata needed for achieving five layers of interoperability: 1) Information Technology (IT) systems must work with other IT systems in implementation or access without any restrictions or with controlled access for technical interoperability; 2) common data formats and communication protocols for syntactic interoperability; 3) contextual semantics related to common semantic resources for semantic interoperability; 4) contextual processes related to common process resources for organizational interoperability; and 5) contextual licenses related to common license resources for legal interoperability (15). Respective metadata can itself form FAIR Digital Objects that can be linked using their persistent identifiers. Consequently, FAIR Digital Objects can exist at different levels of granularity, and coarser level FAIR Digital Objects may consist of several finer level FAIR Digital Objects.

The concept of a semantic unit can be adapted to that of FAIR Digital Objects. Each semantic unit possesses its own UPRI and thus already provides the required persistent identifier. Further, they can be accessed and searched using common protocols such as SPARQL and CYPHER implemented in web interfaces using HTML, with RDF, JSON, and other formats as well-supported data export formats. If a knowledge graph uses well-established controlled vocabularies and ontology terms and organizes its data and metadata following established graph-patterns using tools such as SHACL (32), ShEx (59,60), or OTTR (61,62), all data in the data graphs of semantic units will be semantically interoperable. Additionally, each semantic unit can have its own copyright license assigned, which can either be provided automatically by the knowledge graph application or individually by the creator of a particular semantic unit.

Through additional triples, relevant provenance data can be tracked about each particular semantic unit in the form of either corresponding property-value pairs (property graph) or through an additional provenance Named Graph (RDF/OWL) using well-established provenance vocabularies. The **provenance metadata of a semantic unit** can cover, e.g., creator, creation date, creation application, title of the semantic unit, all contributing users, and last-updated date and are restricted to provenance data about the semantic unit itself—they do not refer to the original process of data production and thus the provenance of the data contents of this unit.

Additionally, if the knowledge graph application makes a distinction between published and unpublished data, **publication metadata** can be tracked separately, covering the date the semantic unit has been published together with its creator and the list of its contributors.

**Access control metadata** specifies any copyright licenses as well as, in case the knowledge graph application differentiates between different groups or roles of users, a specification of who is allowed to access the semantic unit.

With respect to the FAIRness of data, it is also advisable to provide **modelling metadata**, i.e., information about which data schema or graph pattern has been used for modelling the data in the unit's data graph. Each instance of a given semantic unit class should ideally apply the same graph pattern. For this purpose, it would be good to have schemata and patterns, each with its own UPRI, stored in some open repository or having the schemata and patterns be documented in the corresponding semantic unit ontology class.

**Provenance metadata referring to the process of data production** (see, e.g., Fig. 4) can be documented in the form of metadata statement units and metadata item and item group units. The provenance of the data production is usually more detailed and complex than the unit's own provenance and must be documented as semantic units and thus FAIR Digital Objects in their own right.

## Implementation

### Semantic Units, Nanopublications, and Micropublications

The concept of assertional statement units as FAIR Digital Objects is very similar to that of nanopublications—especially, when implementing them in RDF/OWL. **Nanopublications** are RDF graphs that represent the smallest units of publishable information extracted from literature and enriched with provenance and attribution information (63–66). They utilize Named Graphs and Semantic Web technologies. Each nanopublication models a particular assertion, e.g., a scientific claim, in machine-readable format and semantics and is accessible and citable through a unique identifier. Nanopublications are used to facilitate the discovery, exploration, and re-use of scholarly assertions (66). Moreover, embedded in a decentralized network of services that employ semantic templates, nanopublications can also be used by users to directly publish small Linked Data statements as nanopublications (65).

Each nanopublication is organized into four Named Graphs:

1. the *head* Named Graph, connecting the other three Named Graphs to the nanopublication's unique identifier;
2. the *assertion* Named Graph, containing the assertion modelled as a graph;

   3. the *provenance* Named Graph, containing metadata about the assertion; and

   4. the *publicationInfo* Named Graph, containing metadata about the nanopublication itself.

Nanopublications are a natural fit for representing and publishing assertional statement units, with the *assertion* Named Graph closely corresponding to the data graph and the *head* Named graph to the semantic-units graph of an assertional statement unit. Triples in the *provenance* Named Graph can potentially link to other semantic units and thus other nanopublications that contain detailed metadata descriptions (e.g., a graph as shown in Fig. 4). Nanopublications are also suited for publishing the other types of statement units following the same schema. The compound units, on the other hand, correspond to annotated sets of nanopublications, where these sets can either be defined statically (e.g., as nanopublication indexes (67)) or dynamically (by specifying a query on the knowledge graph).

By using dataset units, several such semantic unit nanopublications can be combined to form a micropublication. **Micropublications** (68) represent scientific arguments, in which a specific assertion represents a claim and additional assertions provide supporting or contradicting scientific evidence in the form of data and its accompanying metadata as well as information about attribution, supporting and contradicting references, links to empirical data and figures, etc. Further assertions can be added, pointing to relations to other claims made by other scientific arguments. This way, micropublications relate scientific claims in the form of nanopublications to all relevant information to turn the claim into a scientific argument (*scientific argument* could constitute a specific type of dataset unit).

**Structuring a knowledge graph into interrelated semantic units using the Nanopublications-schema**

As a first step for structuring a knowledge graph into interrelated data graphs, each of which is organized as a particular semantic unit, a primary layer of abstraction above the level of triples must be created. This is achieved by partitioning the knowledge graph into a set of statement units, i.e., a mapping of triples to statement units, where each triple belongs to exactly one data graph of a statement unit.

When implementing this in RDF/OWL, statement units can be modelled as nanopublications (see above). In Neo4j, as it is a property graph, to identify all triples belonging to the same statement unit, one has to add a '*statement_unit_URI:upri*' property-value pair to each of its nodes and relations, with the statement unit's UPRI as the value.

Based on this primary abstraction layer of statement units, a secondary abstraction layer of compound units can be organized. Each compound unit is specified as a collection of two or more semantic units (i.e., statement units and other compound units) by relating the compound unit's UPRI to the UPRIs of its associated semantic units. In RDF/OWL, this can be implemented by using the compound unit's semantic-units graph as the *head* Named Graph of a corresponding nanopublication, while leaving the nanopublication's *assertion* Named Graph empty. The *head* Named Graph thus specifies all basic and compound units that are associated with this compound unit. In contrast, in Neo4j, the nodes and relations of all triples belonging to a compound unit possess a '*compound_unit_URI:upri*' property-value pair with the compound unit's UPRI as their value. Since a given statement unit can be associated with more than one compound unit, its '*compound_unit_URI*' property can have an array of UPRIs of different semantic units as its value.

## Modularity results in increased flexibility of data management

*"Because human thoughts are combinatorial (simple parts combine) and recursive (parts can be embedded within parts), breathtaking expanses of knowledge can be explored with a finite inventory of mental tools."* ((69), p.360).

Just like human thought, structuring a knowledge graph into separate but recursive subgraphs, each of which belongs to a particular semantic unit that, in turn, is classified into different semantic unit classes, organizes the knowledge graph into **structured modules** that can be **(recursively) combined** and that **encapsulate complexity into manageable units** which significantly increases the graph's expressivity and the flexibility of data management. By allowing the specification of relations between semantic units, a given semantic unit from a finer level of representational granularity can be associated with more than one unit from a coarser level. A given statement unit can thus be associated with more than one compound unit, while having the statement unit itself with all its triples being documented in the graph at a single location.

For example, the description of a multicellular organism could contain a has-part statement unit that states that the organism has a head as its part. This statement unit would be associated with the item unit of the organism itself, which would be linked to further item units about the organism's other parts, constituting an item group unit. Moreover, since has-part is a partial order relation (70), the has-part statement unit would also be associated with a parthood granularity tree unit and its corresponding granular item group unit. Thus, the has-part statement unit would be associated with at least four other semantic units.

Due to the modular organization of semantic units, knowledge graphs that implement them can be built as **virtual federated knowledge graphs**, with data being contributed from several projects and institutions, stored in their own repositories. This way, users with access to all data can utilize the contents of such a knowledge graph, while data stewardship remains in the hands of the individual contributing projects and institutions, ensuring their technical autonomy.

The content modularization brought about by semantic units could also facilitate partitioned-based querying of knowledge graphs. This has been successfully done with other approaches for partitioning the graph (71), and we expect that partitioning the graph into semantic units and thus semantically meaningful subgraphs should have a comparable effect on querying tasks. However, the use of semantic units in knowledge graph partitioning and modularity is subject to future research.

# Organizing a knowledge graph into five levels of representational granularity, into other granularity trees, and into different frames of reference using semantic units

With statement unit, item unit, and item group unit as different categories of semantic unit, a knowledge graph can be organized into the following basic representational granularity levels (Fig. 14): triples, statement units, item units, item group units, and the knowledge graph as a whole. Triples represent the lowest and thus finest level of abstraction possible in a knowledge graph, with semantic units constituting coarser levels of abstraction and thus coarser levels of representational granularity. **Representational granularity organizes the semantic-units graph layer and thus the discursive layer of a knowledge graph into different levels of representational granularity**, and with that, indirectly, also the knowledge graph's data graph layer.

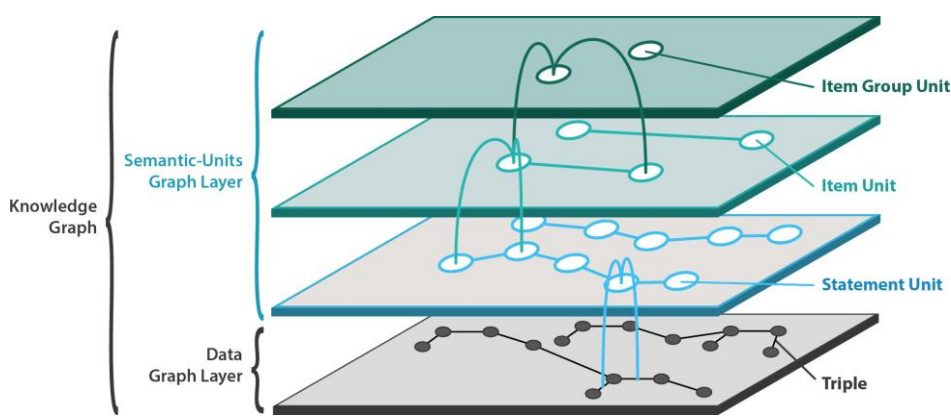

**Figure 14: Five levels of representational granularity.** The introduction of semantic units to a knowledge graph adds a semantic-units graph layer to its data graph layer, which adds, among others, a level of statement units, a level of item units, and a level of item group units to the level of triples and the level of the graph as a whole, resulting in five levels of representational granularity.

Organizing triples into statement units (which represent the smallest units of semantically meaningful propositions for a human reader), statement units into item units (which can be presented in the UI as individual pages), and item units into item group units (which can be presented as collections of semantically interrelated UI pages) is often necessary to increase the human-readability and usability of a knowledge graph.

In addition to the organization and structuring of a knowledge graph into different levels of representational granularity, a given item group unit can be further organized and structured in a somewhat orthogonal way into granular item group units by utilizing granularity tree units present in the item group unit. **Granularity trees organize the data graph layer and thus the ontological layer of a knowledge graph into different granularity perspectives**.

Finally, a given item group unit can be further organized and structured into different context units, each of which covers a specific frame of reference. Points of contact between different frames of reference and thus different context units can be identified via is-about statements: the subject of an is-about statement is a semantic unit resource whose subject belongs to the data graph of one context unit and the object of the is-about statement belongs to the data graph of another context unit (see also Fig. 16). This allows traversing different frames of reference in a knowledge graph via their is-about statement connections. **Context units organize the data graph layer and thus the ontological layer of a knowledge graph into different frames of reference**.

Organizing data into various kinds of semantic units and thus semantically meaningfully subgraphs that can be digested by a human reader as semantically related larger chunks of data is an effective way of presenting data in the UI. With the granular organization of a knowledge graph into different types of semantic units, some of which belong to different levels of representational granularity, and by structuring the knowledge graph into these partially overlapping, partially enclosed subgraphs, data management, data access, data exploration, and graph navigation can be improved significantly.

# Statements about statements and documenting ontological, diagnostic, and discursive information in knowledge graphs using semantic units

Because each semantic unit has its own UPRI and is represented in the semantic-units graph layer of a knowledge graph by its own node, it is straightforward to make **assertions about semantic units** and thus **statements about statements**. If semantic units were organized as FAIR Digital Objects, one could also make statements across different databases and knowledge graphs. By structuring the knowledge graph into semantic units, UIs can be developed that provide a very intuitively usable framework for making statements about statements. The development of such UIs and their accompanying tools and services thereby benefits from semantic units being implemented following a clear and straightforward schema. This way, semantic units can contribute to a solution of the problem that making statements about statements within knowledge graphs can be challenging (see *Challenge 3*). There is another aspect in knowledge management and knowledge representation, where it is beneficial to be able to make statements about statements in a straightforward way and to have a formalized notation for representing contingent statements.

As a scientific realist, one can distinguish between (1) particulars and universals as mind-independent **real entities**, (2) our **cognitive representations of real entities** in the form of ideas, concepts, and thoughts, and (3) **textual and perception-based representations of real entities** in the form of words, symbols, images, recordings, etc. (72–74). The latter are usually used for communication and documentation, with the sender having the intention to trigger cognitive representations in the receiver that are maximally similar to their own representations.

**Meaning** and **reference** are key for the success of communication and condition the duality of a textual representation that must provide both ontological and diagnostic information. The information sent must carry **semantic meaning** in the sense that it provides answers to questions of the type "*What is it?*" and thus **ontological definitions** that carry semantic conceptual content for all referenced entities. Some of this ontological information is provided explicitly, but in most parts the sender relies on the inferential lexical competence (75) of the receiver by using well-known terminology.

But communication also has to provide answers to questions of the type "*How does it look?*" and thus information about the epistemological appearance—information that is required for **unambiguously referencing** all entities mentioned in the text. This **diagnostic**, i.e., **referential information** may not be very important in cases where the receiver already knows the entity, but in all other cases, especially when dealing with theoretical entities whose existence cannot be identified/observed directly and where you often cannot deduce recognition criteria from the ontological definition (e.g., a lead atom, a cell nucleus), diagnostic information is essential for reliably referencing. Therefore, some of this diagnostic information needs to be provided explicitly, but in most parts the sender relies on the referential lexical competence (75) of the receiver.

How does that relate to knowledge graphs? Well, sometimes we talk about real entities, sometimes we talk about our cognitive and textual representations of them. Sometimes, this discourse overlaps, and this is where it becomes complicated when having to model this in a knowledge graph. Ideally, ontological definitions and diagnostic operational definitions are provided by domain ontologies, and assertions in knowledge graphs refer to respective ontology terms and the information linked to them. This way, knowledge graphs would efficiently document, represent, and communicate empirical data.

Unfortunately, however, due to limitations of OWL and Description Logics in defining classes, ontologies generally do not provide diagnostic operational definitions. Ontologies define classes in the form of Aristotelian definitions (76) that constitute essentialistic classes. **Essentialistic classes** are based on **universal statements**, with class membership being determined in dependence of a set of properties that are **individually necessary** and **jointly sufficient** (77). Many diagnostic operational definitions, however, provide method-dependent recognition criteria that require the manipulation of the object to be able to determine its class membership, for instance because the object is too small for the eye, or it can only be identified after it has been stained or when using some test-kit. The defining properties of such operational definitions often refer to properties that are not necessarily present in every instance of a respective class. Instead, they are rather typical and only sufficient in a specific number. Consequently, many recognition criteria cannot be modelled as essentialistic classes but, instead, have to be modelled as cluster classes (74). Contrary to essentialistic classes, an instance of a **cluster class** does not have to possess all class-defining properties—necessary is that some of them apply, sufficient is a specified percentage threshold, which is the **minimum quorum** (77). Definitions of cluster classes thus take the form of **contingent statements**, which cannot be represented in OWL and Description Logics (see discussion above).

Diagnostic operational definitions frequently depend not only on cluster class concepts but also on fuzzy sets, since many cluster class definitions rely on terms that refer to **fuzzy sets** (74). For instance, when recognition criteria refer to *coarse inner material*, *electron-dense clusters*, *mushroom-shaped*, *ovoid*, or *hooked* to identify class-membership of instances to a specific type of anatomical structure, they refer to terms that specify spatio-structural properties that are better communicated via exemplars or images and thus perceptual contents than textual definitions. Whereas both essentialistic classes and cluster classes are always defined textually, fuzzy sets can be defined ostensively in reference to exemplars and thus to perception-based contents that rely on the human capabilities of pattern recognition.

The semantic-units approach provides a formal notation for representing contingent statements and, when combined with semantically annotated exemplary images, a knowledge graph could provide all diagnostic knowledge relevant for answering *"How does it look?"* questions. Together with the ontological definitions, knowledge graphs could provide any reader of the graph with all relevant knowledge they need to ensure their inferential and referential lexical competence.

In addition to the ontological and the diagnostic layer, knowledge graphs often document a third layer of information, representing **discursive contextual information** (see also Ingvar Johannson's distinction between *use* and *mention* of linguistic entities (78)). In other words, knowledge graphs not only contain empirical data in the form of assertions like "*The melting point of lead is at 327.5 °C*", but also assertions like "*author A asserts that the melting point of lead is at 327.5 °C*" or "*the assertion that the melting point of lead is at 327.5 °C is a result of experiment X*", which address the discursive contextual information. Some assertions may even address the diagnostic aspects of how the reference between textual representation and real entity is achieved by documenting respective methods and diagnostic criteria. Without a good method for making statements about statements, documenting and representing these three layers of information (i.e., **ontological**, **diagnostic**, **discursive**) and expressing their interrelations is not possible.

Because semantic units are represented in the graph with their own nodes, they support the documentation and representation of all three layers and their interrelationships. Statement units can provide within their data graphs **ontological information** that can be gathered within compound units of coarser representational granularity. As propositions, they are represented in their semantic-units graphs, which constitutes a large part of the **discursive layer**. Moreover, contrary to OWL and Description Logics, the semantic-units approach with its introduction of some-instance and every-instance resources and resources that represent individual universal statements and sets of semantically related universal statements, enables differentiated referencing of various entities relating to the ontological layer. Consequently, one can make statements about individual classes (=concepts), their definitions, their individual defining statements, a single instance of a class, a group of instances of a class, and the group of all instances of a class, therewith opening multiple ways to talk about universal statements and their elements. Moreover, with context units, it is even possible to distinguish and explicitly document **different frames of reference** within the ontological and discursive layer. In other words, the semantic-units approach allows for specifying various interrelationships between the discursive and the ontological layer of its knowledge graph and therefore integrates universal statements in the knowledge graph's domain of discourse (*Challenge 4*).

Since statement units can also relate a particular resource to another statement unit (e.g., '*author_A -asserts-> statement_unit_Y*') or two statement units to each other (e.g., '*statement_unit_X -hasMetadata-> statement_unit_Z*'), relations between the three layers can be easily documented. For instance, an item group unit documents the contents of a particular scholarly publication. Because scholarly publications can be considered in general to represent reports about a scientific investigation or some other research activity, the item group unit will contain information about a research activity as a process that may have some material entities (i.e., substances, instruments, specimens, etc.) and research agents (i.e., experimenters, patients, test persons, etc.) or some data as its input and some research result as its output. Moreover, the research activity usually realizes a specific plan (i.e., research method) and achieves a specific research objective, which is part of the plan (see Fig. 15). This general model can be applied like a Matryoshka, the russian stacking doll: research activities can have research activities as their parts, i.e., differentiated steps in the overall research process, that again have their own inputs and outputs and realize their own research methods and achieve their own objectives.

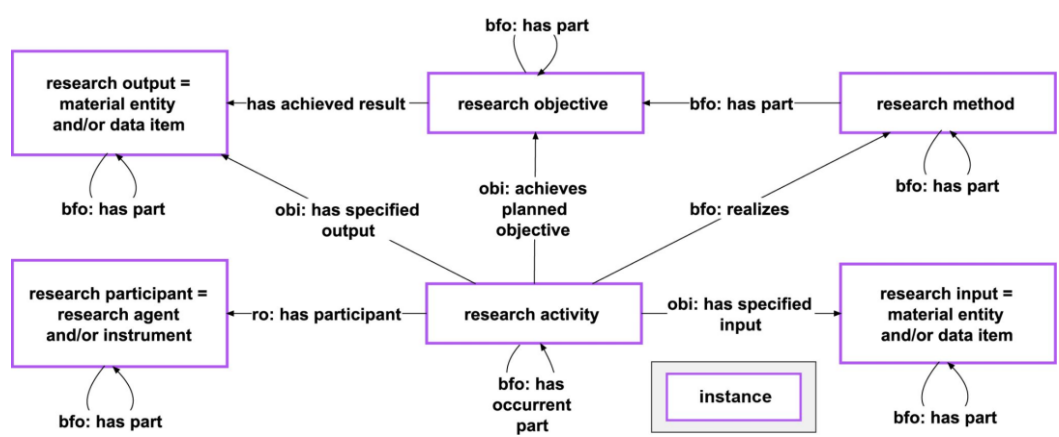

**Figure 15: A data scheme for modelling the contents of scholarly publications.** The relation between a research activity, its input and output and its underlying process plan specification in the form of a research method and research objective (model adapted from (79)).

This model can be applied to a knowledge graph of scholarly publications and be adopted to semantic units. The item group unit of the publication is about an instance of *research activity* and has an item unit of the research activity associated with it, of which that *research activity* instance is the subject. The research activity item unit has a research result item unit associated with it that has an instance of *research result* as its subject. The research activity resource is linked to the research result resource via the property *hasSpecifiedOutput* (OBI:0000299). Now, the research result item unit can be linked to a description item unit that contains several assertional statement units about a particular multicellular organism—the publication describes the anatomy of a specific multicellular organism. This is documented through a triple that relates the *research result item unit* to the *description item unit* via the property *hasLinkedSemanticUnit*. The 'description' (SIO:000136) resource is the subject of the *description item unit* and is related to the multicellular organism item unit through the property *hasPart* (BFO:0000051) and the multicellular organism item unit to the instance of multicellular organism (UBERON:0000468) through the property *hasSemanticUnitSubject*. The resulting set of semantic units thereby cover at least three different frames of reference: the publication itself, the research activity, which the publication is reporting on, and the assertional statements about the research subject, i.e., the multicellular organism, with their boundaries being indicated by the two is-about statements. In Figure 16, you can see how the resulting discursive and the ontological layers, as well as the different frames of reference, are interconnected.

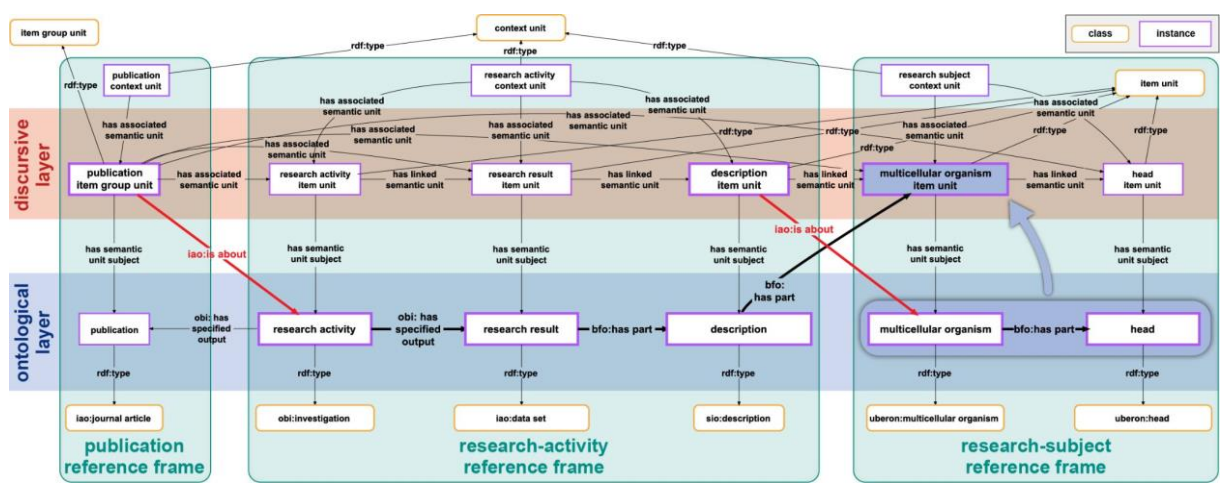

**Figure 16: Detail from the graph belonging to an item group unit about the contents of a scholarly publication, represented in an ABox RDF graph.** The content is modelled according to a specific data schema (see Fig. 15) that has been adapted to semantic units. All content from the published journal article is organized in its own item group unit instance via various associated semantic units. The publication itself is represented in the data graph as an instance of journal article (IAO:0000013). The item group unit has several item units about the research activity associated with it, that in turn are connected to each other via the property *hasLinkedSemanticUnit*. The item unit describing the research activity has an instance of investigation (OBI:0000066) as its subject, which has as its output an instance of data set (IAO:0000100) that has an instance of description (SIO:000136) as its part. That description instance has the item unit that describes the multicellular organism as its part, the latter of which has an instance of multicellular organism (UBERON:0000468) as its subject. The blue arrow indicates that the data graph (dark blue box without borders) is represented by this item unit (bordered box in the same color). It has a head item unit linked to it, which has an instance of head (UBERON:0000033) as its subject, which is a part of the multicellular organism. The data graphs of the semantic units form the ontological layer, whereas their semantic-units graphs form the discursive layer. Three different context units separate the reference frame of a publication from that of a research-activity and that of a research-subject, with is-about statements marking their borders. *For reasons of clarity of presentation, the associated statement units are not shown in the discursive layer.*

Whereas the resources of the semantic units represent and organize the discursive layer, their data graphs represent the ontological (and diagnostic) layer. The same applies to all semantic unit resources. The organization of a knowledge graph into these different layers and into different frames of reference may provide a new framework for the development of innovative visualization and graph exploration methods.

# Modelling negations, cardinality restrictions, and disagreement

OWL and Description Logics based information systems adhere to what is called the **Open World Assumption** (OWA). OWA assumes incomplete information by default. Therefore, absence of information about an entity or a fact does not necessarily imply information about the absence of that entity or the negation of that fact. As a consequence, if you for instance want to state that a particular object is a fruit but not a pome fruit, that a particular head of an insect does not possess any antenna, or that this head possesses exactly three eyes, it is not enough to mention that the object is a fruit, the head possesses three eyes, or to not mention any antenna, because due to OWA the fruit could still be a pome fruit and the head could still possess antenna, and it could still possess a fourth eye. Instead, you must explicitly state all these things.

Unfortunately, in RDF/OWL, these types of negation and cardinality statements pose tricky modelling challenges, because they cannot be directly expressed as relations between instances and thus ABox expressions. Instead, you must model them in the form of class-expressions as a TBox, e.g., by using OWL Manchester Syntax (51).

Regarding **negations**, one can distinguish two different types of negation statements: negations involving references to classes and negations of relations between instances. An example of the former is the statement '*this fruit is not a pome fruit*' (Fig. 17A). This can be characterized following the Manchester Syntax as the expression 'not (type pome fruit)' (Fig. 17B), which translates to OWL mapped to RDF to a graph in which the given particular entity is an instance of the class fruit (PO:0009001) and an instance of a class that is the complement of the class pome fruit (PO:0030110) (Fig. 17C). Analog to this case are negations that are part of a class axiom.

Semantic units can provide an alternative modelling solution by introducing a general *negation unit* class (in Neo4j this could be simply modelled by adding the label '*:Negation*' to the respective semantic unit node, indicating that this resource is an instance of the class 'negation'). The statement '*this fruit is not a pome fruit*' can be modelled by two semantic units, both of which would be instances of *named-individual identification unit* and *assertional statement unit*. However, the semantic unit stating that its subject is a pome fruit (PO:0030110) is also an instance of *negation unit*, thereby negating the contents of its data graph (see Fig. 17D).

**Figure 17: Modelling negations involving instances by applying semantic units. A)** A human-readable statement that this fruit is not a pome fruit. The statement can be modelled in two different ways. As an OWL expression that can be specified using **B)** Manchester Syntax, where the fruit is an instance of a class that is defined as all of its instances are not instances of pome fruit (PO:0030110). Note, how this Manchester Syntax expression translates into **C)** an OWL expression mapped to RDF, where fruit x is an instance of fruit (PO:0009001) but also of a class that is the complement to pome fruit (PO:0030110). Alternatively, the statement can be modelled as an instance-based graph using semantic units **D)**. The data graph in the blue box states that the entity is a fruit. The data graph belonging to the

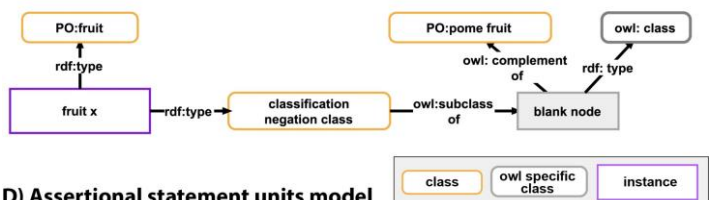
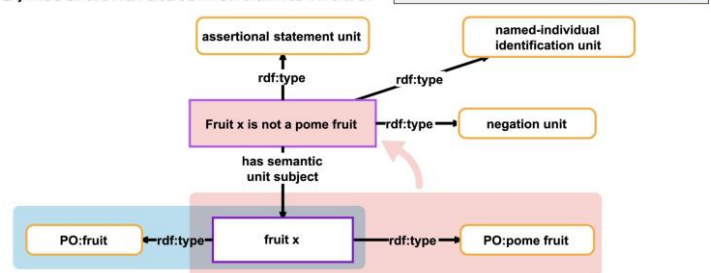

statement unit stating the negation (red box with borders, here shown with its dynamic label), on the other hand, is shown in the red box without borders and states that the entity is a pome fruit. However, since the latter statement unit not only instantiates the classes *assertional statement unit* and *named-individual identification unit* but also *negation unit*, it actually negates the statement in the data graph, therewith indicating that it is *not* a pome fruit.

Absence statements are another example of negations involving reference to classes. The observation '*this head has no antenna*' translates to the statement that the head is an instance of a class that is characterized following the Manchester expression as 'not ('has part' some antenna)', which translates to OWL to a more complex class axiom (see Fig. 18A-C).

The same statement can be modelled as the union[6] of the subgraphs of two semantic units, with the one having the 'head x' (UBERON:0000033) as its subject being an instance of *has-part statement unit*, *assertional statement unit*, and *negation unit*, whereas the one having the 'some antenna' (UBERON:0000972) as its subject only being an instance of *some-instance identification unit* (see Fig. 18D). In general, when modelling absence statements this way, all instance resources in the assertion that are supposed to represent types of entities that are absent, would be modelled this way, relating to the respective class via the property *some instance of* within their corresponding some-instance identification units. This notation would be simpler and easier to use and to implement in knowledge graph applications than the OWL-based notation of using a more complex class expression.

---

[6] The union of the subgraphs, but, semantically, the intersection of the statements.

**Figure 18: Relation between an absence observation, the corresponding assertions, and two alternative ways to model them in a knowledge graph**. **A)** A human-readable statement about the observation that a given head has no antenna. **B)** Absence statements cannot be expressed as relations between instances. Therefore, the observation from A) must be expressed using a class expression, which can be formulated using Manchester syntax. Following this notation, the head would be an instance of a class that is defined to have only instances that have no antenna as their parts ('not' and 'some' being used as mathematical expressions). **C)** The translation of the assertion from A) and B) into an OWL expression mapped to RDF. Note how *absence phenotype* is defined as a set of relations of subclass and complement restrictions involving two blank nodes. **D)** The same statement can be modelled using two semantic units. One of them is modelling the has-part relation and negates it (red box with borders and its dynamic label, with its data graph in the red box without borders). It is therefore an instance of *has-part statement unit* as well as *assertional statement unit* and *negation unit*. The other semantic unit is an instance of *some-instance identification unit* and relates 'some antenna' to antenna (UBERON:0000972) via the property *some instance of*. Its data graph is shown in the blue box. Together, they model the observation from A). This notation is comparable to E). **E)** An alternative notation of the statement 'this head has no antenna' and the observation from A). The notation uses Peirce's predicate logic system of existential graphs. The *identity line* ⎯ between the two phrases '*head x*' and '*has part antenna*' states that head x has some antenna as its part, whereas the red circle surrounding the latter phrase expresses its negation by crossing the *line of identity*. *For reason of clarity of representation, the relation between 'head x' and head (UBERON:0000033) is not shown in C) and D).*

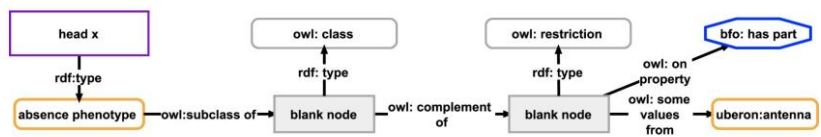

**A) Observation**
*This head has no antenna*   (negated parthood assertion)

**B) OWL Manchester Syntax expression**
absence phenotype class axiom:    not ( *has part* some antenna )

**C) Conventional OWL model**

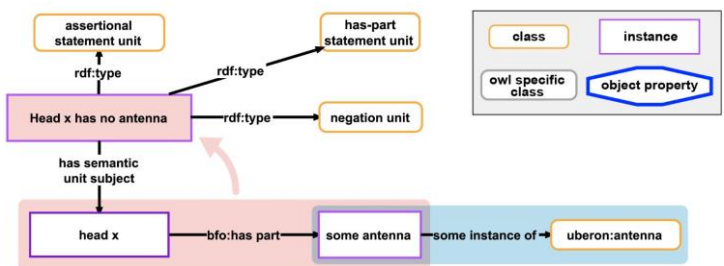

**D) Assertional statement units model**

**E) Peirce's predicate logic system of existential graphs**

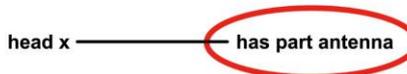

The same approach can also be applied to the case of negations of relations between two instances. The fact that a given fruit is not part of a particular orange plant could be modelled analog to the other negation statements (see Fig. 19).

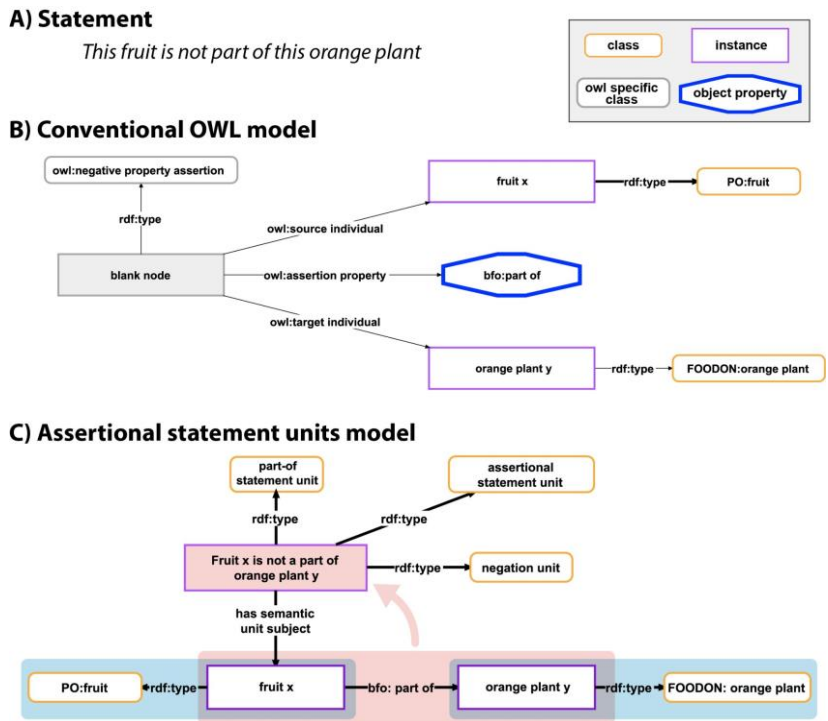

**Figure 19: Modelling negating relations between instances by applying semantic units**. **A)** A human-readable statement that this fruit is not part of this orange plant. The statement can be modelled in two different ways: **B)** as an OWL expression mapped to RDF. Note, how the statement is translated into a negated assertion statement with source, property, and target specification, relating an instance 'fruit x' (PO:0009001) to an instance 'orange plant y' (FOODON:03411339) involving a blank node; or **C)** using three semantic units. The two named-individual identification units, with their data graphs shown in the blue boxes, state that one of the objects is a fruit and the other one an orange plant, whereas the part-of statement unit (red bordered box and its dynamic label, with its data graph shown in the red box without borders) states that fruit x is not part of orange plant y as it instantiates both *part-of statement unit*, *assertional statement unit*, and *negation unit*.

From all these examples, it becomes clear that negations, including absence statements, can be efficiently represented via semantic units by classifying the corresponding semantic unit as a *negation unit*. This general notation can be compared to Peirce's existential relational graphs (80,81), where a *line of identity* '—' can be used to express that 'something is A' by notating it as '—A'. The *line of identity* can be understood to represent an existential quantifier ($\exists x$). By interrupting this line with a circle that encloses A, you express that 'something is not A' (see Fig. 18E). The resulting existential relational graphs are sufficiently general to represent full first-order logic with equality (81).

Expressing **cardinality restrictions** is, for the same reasons, also challenging using RDF/OWL and requires the use of a class expression that indicates the cardinality as a class restriction. Instead of using a rather complex class axiom to describe for instance that a given head possesses exactly three eyes (Fig. 20B), the same information could be modelled by two semantic units and allowing for extending the some-instance identification unit to include cardinality restrictions via linking its subject to, for instance, the value 3 via the property *qualified cardinality* (OWL:qualifiedCardinality) (Fig. 20C). Instead of a simple integer value, a float value range with a unit specification that is constrained to count unit (UO:0000189) and percent (UO:0000187) would even allow the specification of frequencies and ranges. The has-part statement unit relates the 'head x' to the subject of a semantic unit that instantiates both *some-instance identification unit* and *cardinality restriction unit*.

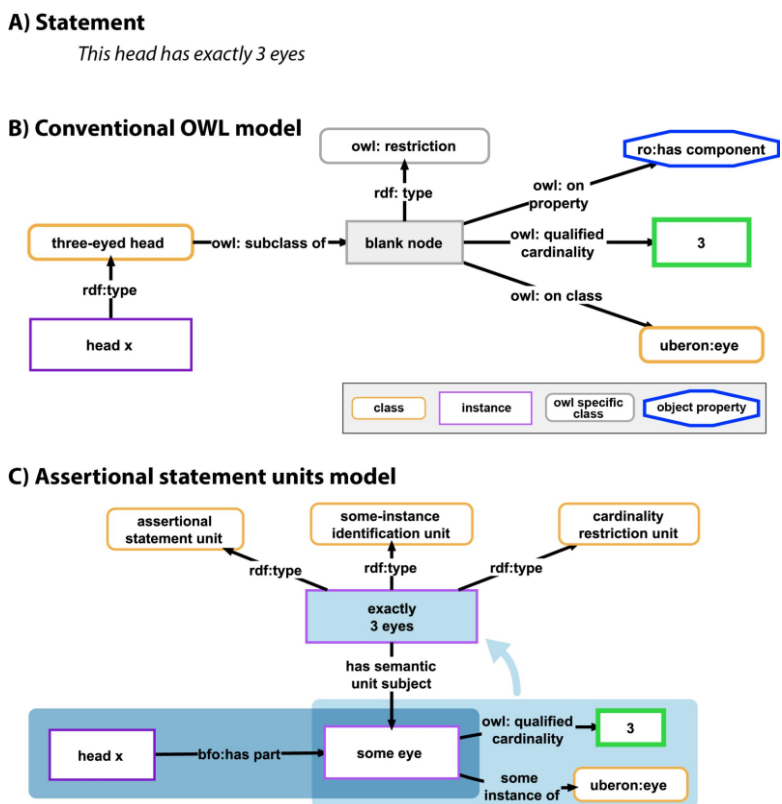

**Figure 20: Modelling cardinality restrictions involving instances by applying semantic units**. **A)** A human-readable statement that this head has exactly three eyes. The statement can be modelled in two different ways: **B)** as an OWL expression mapped to RDF. Note, how 'has exactly three eyes' is translated into being an instance of a class that has a cardinality restriction on the *has component* property (RO:0002180) and the class *eye* (UBERON:0000970) with a cardinality value of 3, thereby involving one blank node; or **C)** using two semantic units, one of which models the has-part relationship, with its data graph shown in the dark blue box. The other semantic unit (light blue bordered box and its dynamic label, with its data graph shown in the light blue box without borders) instantiates *assertional statement unit*, *some-instance identification unit*, and *cardinality restriction unit*.

Analog to the approach for modelling negations, semantic units can also be applied for modelling **disagreement** in a knowledge graph. For instance, if person A asserts 'This fruit is a pome fruit' through a named-individual identification unit, and person B disagrees with this statement, the disagreement can be modelled as a semantic unit that instantiates both *assertional statement unit* and *disagreement unit* and its data graph states that the named-individual identification unit asserted by person A instantiates *negation unit* (see Fig. 21).

Whereas the here suggested notations result in simpler graphs than their OWL-based RDF equivalents, and these graphs are easier to query due to the lack of blank nodes, reasoners that directly use semantic units would still have to be developed; however, through the translation to OWL (see section Logical semantics of semantic units), standard OWL reasoners will be able to reason over semantic units, including negations, cardinality restrictions, and disagreements.

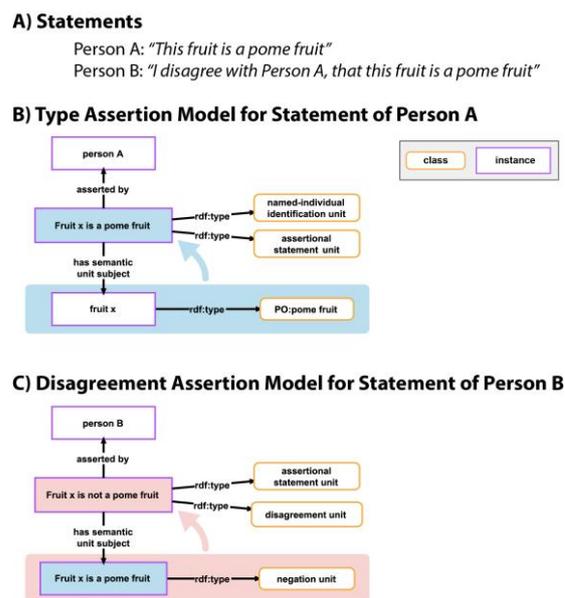

**Figure 21: Modelling disagreement by applying semantic units**. **A)** Person A states that this is a pome fruit and person B disagrees. **B)** The assertional statement unit (blue bordered box and its dynamic label, with its data graph shown in the blue box without borders) models the statement of person A. **C)** The disagreement unit (red bordered box and its dynamic label, with its data graph shown in the red box without borders) models the statement of person B and instantiates *assertional statement unit* and *disagreement unit*. Its data graph states that the statement of person A instantiates *negation unit*.

## Logical semantics of semantic units

Semantically, we treat semantic units as translations between ABox and TBox statements, or, alternatively, as translations between logic programs and OWL axioms. For example, 'Lars' right hand' (FMA:9712) *has-part* (BFO:0000051) 'Lars' right thumb' (FMA:24938) is an ABox statement. It includes, as part of the representation, the information that this expresses a "has part" type axiom (has-part statement unit). Semantically, we treat these statement units as expressions of particular forms of ontology design patterns, where the ABox statement is translated into TBox axioms based on an ontology design pattern dependent on the type of statement unit. We can formalize these translations using expressions of **relational ontology design patterns** (82), which are OWL axioms involving variables for arbitrary entities. For example, the *'has-part assertional statement unit'* can be translated, in general, as

   *?X SubClassOf: has-part some ?Y,*

where *?X* is filled by the subject and *?Y* by the object of the statement. We can add the pattern as literal, i.e., either as a datatype property or annotation property, to the statement unit class. This will allow use of an OWL reasoner, or, alternatively, a SPARQL query, to retrieve all statements that can be rewritten based on this design pattern; however, it will be more convenient to specify them separately and use a dual formalism.

The dual representation of complex axioms as ABox statements and the translation into TBox (or ABox) axioms enables a **dual kind of reasoning**. Through the translation to TBox axioms, it becomes possible to apply OWL reasoning and therefore use any **OWL reasoner** (or reasoners for OWL profiles) to test consistency of the knowledge graph or perform inferences. However, the ABox representation of the axioms as abbreviated patterns also enables a form of reasoning directly on the ABox statements, using the semantics of **logic programming**; this form of reasoning can exceed the semantics of OWL (83). In particular, it becomes possible to add **non-monotonic statements** which allow us to model contingent statements, or "defaults", which may have an exception (84). For example, the statement '*Typically, a hand has a thumb as part*' can be modelled through a logic program that states that, if there is an instance *x* of Hand (FMA:9712) and it is not provable that *x* does not have some instance of Thumb (FMA:24938) as part, then *x* has some instance of Thumb as part. Formally, this can be expressed as an **answer set program** such as

*has-part(x, Thumb) :- rdf:type(x, Hand), not lacks-part(x, Thumb)*.

However, such a statement has several unusual features that need to be explained. First, it is not based on OWL semantics and therefore does not, initially, interact with axioms in the OWL ontology; in particular, the statement unit is not translated into an OWL axiom but rather into a logic rule that is applied to the statement units themselves. Second, under answer set semantics, "not" is weak negation with the intended meaning that *lacks-part(x, Thumb)* is not provable. It therefore encodes a kind of default knowledge and using this statement results in **non-monotonic inference**: adding a new statement, such as *lacks-part(x, Thumb)*, will invalidate the previous inference (*has-part(x, Thumb)*); this is in contrast to the standard semantics of OWL which is monotonic (where adding a new statement will never invalidate an inference). Third, the symbols *Hand* and *Thumb* are used here as symbols for individuals and *instanceOf* is an explicit relation between entities, whereas *Hand* and *Thumb* would be classes in OWL (corresponding to unary predicates) and *instanceOf* is a built-in primitive that relates individuals to classes.

To relate the logic programs to OWL, we assume that the logic program is able to use all OWL entities (class symbols, individual symbols, and relation symbols) as individuals. An atomic formula takes the form P ($a_1$, ..., $a_n$) where P is a predicate symbol and each $a_i$ is either an individual symbol or a variable symbol. A formula is called ground if it contains no variable symbol. Several efficient grounders, such as Clingo (85), as well as answer set solvers such as DLV (86) or clasp (87) are available, and allow computation of fully ground answer sets (or stable models of answer set programs). We treat the predicate symbols in the answer set program as ontology design patterns that are translated into a set of OWL axioms. Applying patterns in a forward manner is decidable and can be implemented in a straightforward manner by computing the fully grounded stable models of the answer set program and then applying the forward translation of the design pattern. We can also implement an inverse translation from OWL into these patterns by querying, for each defined ontology design pattern, whether the pattern is entailed by the OWL axioms; using an OWL reasoner, we can query for the pattern within the deductive closure $O^\vdash$ of ontology O and iterating through all OWL entities. However, for a pattern with n variables and |E| entities in the OWL ontology, the complexity will be $O(|E|^n)$. Using this approach, a semantic unit is specified by two parts, a statement that takes the form of a logic program, and a pattern to convert predicates into OWL statements. For example, we can express a named-individual identification unit (Figure 8) as:

Logic program:   *rdf:type(Lars'RightHand, fma:hand)*

Pattern:   *rdf:type(Lars'RightHand, fma:hand)*

In the case of a named-individual identification unit, both statements are identical. More generally, we can formulate all named-individual identification units using a logic program such as:

*named-individual-identification-unit(x), has-semantic-subject(x,y), rdf:type(y, z), rdfs:label(y, l), owl:named_individual(y)*;

and a pattern that translates each predicate into their corresponding OWL statement.

A some-instance identification unit is semantically identical to a named-individual identification unit but does not introduce, explicitly, a name for the individual. This corresponds to the use of an existential quantifier which we can eliminate through "Skolemization", i.e., we can introduce a new individual name that is not explicitly referred to in the language of the semantic units but exists in the semantic space.

The every-instance identification unit requires more elaborate translation into OWL. Here, *<u>every instance of X</u>*  refers to the collection of all instances of X, and we may rely on a theory of collections and collectives (88); the example in Figure 8 will then be translated into:

- *rdf:type(everyHand, Collection)*

- *owl:SubClassOf(fma:Hand, owl:SomeValuesFrom(member-of, owl:oneOf({everyHand})))*

- *owl:SubClassOf(owl:oneOf(everyHand), owl:AllValuesFrom(has-member, fma:Hand))*

In other words, *<u>'everyHand'</u>*  refers to a collection that has as its members all instances of the class <u>Hand</u> (FMA:9712), and that has only instances of that class as members. Analogously, we can translate cardinality restrictions (cardinality restriction units); the example in Figure 20, for example, would be expressed using the following (general) logic program:

*cardinality-restriction-unit(x),   has-semantic-unit-subject(x,y),   owl:qualified-cardinality(y,z), some-instance-of(y, w).*

with a translation of this pattern into OWL of

   *rdf:type(cX, owl:intersectionOf(Collection, owl:cardinality(has-member, 3, uberon:eye)))*

where *cX* is a new individual name (i.e., a name not used elsewhere). This will then be combined with the 'part-of assertional statement unit' for 'head X' in the same example to assert the ABox statement

   *part-of(`head X', cX).*

We can use a similar translation to formalize other statements. An assertional statement unit (Fig. 9) is translated directly into an ABox statement, and a universal statement unit follows a similar translation; the statement in Figure 9 will be translated into the OWL ABox axiom

   *hasPart(everyHand, someThumb).*

Based on OWL semantics and the axioms to formalize *'everyHand'* and *'someThumb'* gives rise to the inference of the OWL axiom

   *owl: SubClassOf(fma:Hand, owl:SomeValuesFrom(has-part, Thumb)*

as expected. The contingent statement unit in Figure 9 is similarly translated into an OWL ABox axiom. However, our formalism also allows us to express a different type of contingent statement, i.e., a statement about prototypes (prototypical contingent statement unit). Such a statement can be used to formulate expressions such as '*normally, a hand has a thumb as its part*', allowing for the possibility of an exception (such as in the case of loss of a thumb due to an accident). The following statement expresses this prototypical statement:

   *hasPart(x, thumb) :- hasPart(hand, thumb), instanceOf(x, hand), not -hasPart(x, thumb).*

The use of the logic programming paradigm also allows us to formalize **complex statements** such as shown in Figure 13 which are **not directly expressible in OWL**, and the translation into OWL through translation patterns enables inferences within OWL when combined with other OWL axioms.

We are also able to model (classical) **negation** through the use of negation units (Figure 17). A negation is directly added to a semantic unit, and we can use this to define a negated assertional statement unit:

   *NegatedAssertionalStatementUnit(x) :- NegationUnit(x), AssertionalStatementUnit(x)*

and then a translation into OWL for *NegatedAssertionalStatementUnit* with the following logic program as precondition:

*NegatedAssertionalStatementUnit(x), has-semantic-unit-subject(x,y), rdf:type(y,z)*

and the OWL translation

*rdf:type(y, owl:complementOf(z)).*

This example shows the flexibility of our use of logic programming, as the *NegatedAssertionalStatementUnit* predicate is inferred from the two assertions added to the statement unit and does not have to be asserted. However, this kind of inference, in particular when using negation units, leads to one additional challenge because multiple patterns may apply to statements, and it may not always be clear which translation pattern to apply. For example, in the example in Figure 17, if we ignore the negation unit, we will conclude that the pattern for assertional statement units should apply; applying translations for both statement units will result in an inconsistency. The easiest way to prevent this is to add the weakly negated condition to the logic programs for assertional statement units: *not NegatedAssertionalStatementUnit(x)*, or, even simpler, *not NegationUnit(x)*; this condition will prevent the translation pattern from being triggered in the presence of a negation unit.

Furthermore, the logic programming paradigm can be used to translate between statements (treated as individuals) and their content (treated as propositions). The statements, as shown in Figure 21, are units in their own right (assertional statement units) which can be translated to OWL axioms using the patterns we already described. Additionally, the statements are individuals within logic programs where relations between statements can be modelled, which additionally can enable the use of answer set programming to represent arguments (89,90).

# Semantic units provide a new framework for knowledge graph alignment

Semantic units that instantiate the same semantic unit ontology class contain semantically similar information. Therefore, semantic units of the same type are in principle comparable throughout all their instances. Consequently, aligning and comparing different knowledge graphs that are based on the same set of semantic unit ontology classes can be conducted in a **step-by-step procedure** along the different levels of representational granularity. In a first step, the data graphs belonging to item group units are aligned based on their *_hasSemanticUnitSubject_** relations to different types of subjects and based on their *_hasAssociatedSemanticUnit_** relations to different types of associated semantic units. In a next step, for each pair of item group units the data graphs belonging to their associated item units are aligned based on their types of subjects and their types of associated statement units, which in turn can be aligned by class, and finally their individual triples can be aligned. This would provide a new framework to improve methods for knowledge graph alignment, subgraph-matching, graph comparison, and graph similarity measures.

## Managing restricted access to sensitive data

Statement units and their classification into corresponding ontology classes provide a framework for identifying subgraphs within a knowledge graph that can contain sensitive data to which access must be restricted. For example, all information relating to location data of occurrences of endangered species should not be publicly available, and access should be restricted and managed by adequate rules. Respective statement units can be identified by class, and the need to restrict access to them can be based on the threat level of the respective species. If a species is endangered, access to these types of statement units would be automatically restricted, while all other semantic units referring to that species would still be openly accessible. Similarly, any personal data fall under privacy policy restrictions and access to corresponding statement units should depend on specific access rights. This would follow EOSC's principle of '*as Open as possible, as closed as necessary*' (16).

# Conclusions and Future Work

With semantic units, we introduce a type of resource that is new to knowledge graphs and that significantly **increases their overall expressivity**. Semantic units structure a knowledge graph into identifiable and semantically meaningful subgraphs and thus represent an additional type of **representational entity** (91) besides instances, classes, and properties. **Semantic units represent resources of a higher level of abstraction than the low-level abstraction of the resources used in conventional individual triples**. Technically, semantic unit resources are instances of respective ontology classes, but semantically they represent the contents of their data graphs and thus statements or sets of semantically and ontologically related statements. With semantic units, one can thus **extract semantically meaningful entities from a knowledge graph** that are more abstract than the resources typically used in conventional triples.

Regarding the four challenges discussed in the introduction, we can conclude that because every statement is organized in its own statement unit which, in turn, instantiates a corresponding statement unit ontology class, a **data schema** (e.g., as a shape) and corresponding **CRUD query patterns** can be specified for each such class. By linking the data schema to the corresponding statement unit class, the UPRI of each statement unit and its affiliation with a corresponding statement unit class would also **reference the underlying data schema and thus contribute to the FAIRness of their data by guaranteeing schematic interoperability**. Because statement units partition the knowledge graph so that every triple belongs to exactly one statement unit, all data in the knowledge graph would have information about their underlying data schema. Reference to the underlying data schema represents valuable metadata that guarantees the interoperability of all instances of statement units of the same class because they are all based on the same data schema (*Challenge 1*).

The set of CRUD query patterns would not only provide read queries for domain experts for searching instances of the corresponding class in the overall data graph, but also create, update, and delete queries for developers, who could use them and thus do not have to learn graph query languages anymore (*Challenge 2*).

By organizing the overall data graph into different semantic units and thus semantically meaningful subgraphs at different levels of representational granularity, semantic units also provide a solution to the problem that making **statements about statements** is challenging in knowledge graphs (*Challenge 3*). Semantic units provide an efficient way of structuring the graph to allow users to intuitively make statements about statements, and also provides a clear and straightforward implementation schema that supports the development of corresponding queries and related tools. Moreover, contrary to RDF-star, semantic units also allow distinguishing different instances of the same statement by organizing them as different semantic units that instantiate the same semantic unit class. They would have identical subjects and objects, but would differ by their respective semantic unit resource.

Finally, by distinguishing between assertional, contingent (including prototypical), and universal statement units as top-level categories of statement units, by introducing some-instance and every-instance resources, and by introducing a notation that does not involve blank nodes and that is less complex than the equivalent OWL notation when mapped to RDF, semantic units provide a conceptual framework with which the conceptual gap between RDF/OWL and property graphs can be bridged, formal semantics for contingent and prototypical statements can be provided, and universal statements and thus also particular class axioms can become part of the domain of discourse of a knowledge graph (*Challenge 4*). In other words, **semantic units substantially increase the overall expressivity of knowledge graphs**.

In the future, we want to expand the concept of a semantic unit to include for each semantic unit class corresponding data schemata and query patterns. Together with further features, this results in what we call a **Knowledge Graph Building Block** (KGBB) (see also knowledge graph cells (79) for a first discussion of a similar idea). A KGBB is a small module that is associated with a semantic unit ontology class. Each semantic unit class requires its own associated KGBB. A KGBB provides all information required for managing data belonging to the corresponding type of semantic unit. Each KGBB provides a graph model for storing the data of its associated type of semantic unit, from which CRUD query patterns can be derived, resulting in FAIR data and metadata. KGBBs **decouple data storage from data access** by providing data access in various formats, and they also **decouple data display from data storage** by providing different display templates. Ultimately, the idea for KGBBs is that each KGBB functions independently of other KGBBs, that it is made openly available in a repository so that it can be reused by others, and that domain experts can combine multiple KGBBs and define their possible interactions with as little effort as possible to set up their own knowledge graph application, without being a programmer or having knowledge in semantics. A **KGBB editor** will enable domain experts without background in semantics, in programming, or in graph query languages to describe new KGBBs. A **KGBB application engine** will use the information provided by the various KGBBs of a knowledge graph application and communicate through respective APIs with the persistence-layer and the presentation-layer, thereby **decoupling not only human-readable data display from data storage, but also data access from data storage, and data storage from storage technology**.

One of the authors has developed a first minimum viable product for how KGBBs and semantic units can be used to manage a knowledge graph (92). It is based on Neo4j and gets its contents through a web interface and user input. It is a small knowledge graph application for documenting assertions from scholarly publications and allows users in an exemplary way to describe some contents that can be found in a scholarly publication (it does not focus on describing the publication's bibliographic metadata). Each described paper is represented as its own item group unit, with the assertions covered by statement units that are associated with item units and granularity tree units. The showcase is based on Python and flask/Jinja2 and is openly available through https://github.com/LarsVogt/Knowledge-Graph-Building-Blocks.

We believe that the combination of semantic units with Knowledge Graph Building Blocks will contribute a framework for sharing data across many stakeholders, helping to interlink data providers from a diverse range of different areas, with data stewardship remaining in the hands of the domain experts or institutions, thus ensuring their technical autonomy (following Barend Mons' *data visiting* as opposed to *data sharing* (7)). We also think that due to its modular character, the framework can increase the accessibility of knowledge graphs for software developers and domain experts who lack the expertise in semantics, thus supporting the FAIRification of data and metadata in general—something desperately needed to make all data FAIR, which we believe would increase our chances to fight climate change and biodiversity loss.

# Acknowledgements

We thank Werner Ceusters, Nico Matentzoglu, Manuel Prinz, Marcel Konrad, Philip Strömert, Roman Baum, Björn Quast, Peter Grobe, István Míko, Manfred Jeusfeld, Manolis Koubarakis, Javad Chamanara, and Kheir Eddine for discussing some of the presented ideas. Lars Vogt received funding by the ERC H2020 Project 'ScienceGraph' (819536). We are solely responsible for all the arguments and statements in this paper.

# Author's contributions

LV: developed the concept of semantic units and wrote the manuscript. TC: contributed to the chapter about nanopublications. RH: contributed the formal semantics. All authors: reviewed the manuscript.

*Semantic Units* 64